\documentclass[aps,prd,draft,groupedaddress]{revtex4}
\def\Journal#1#2#3#4{{#1} {\bf#2}, #3 (#4)}

\def\PRD{{\rm Phys. Rev.} D}

\def\ep{\epsilon}
\def\eph{\epsilon_h}
\def\ephp{\epsilon^*_{h^\prime}}
\def\vep{\varepsilon}
\def\la{\langle}
\def\ra{\rangle}

\def\be{\begin{equation}}
\def\ee{\end{equation}}
\def\bea{\begin{eqnarray}}
\def\eea{\end{eqnarray}}
\input{epsfig.sty}
\begin{document}
\title{Vector Anomaly Revisited in the Anomalous Magnetic
Moment of $W^\pm$ Bosons}
\author{ Bernard L. G. Bakker$^{a}$ and Chueng-Ryong Ji$^{b}$\\
$^a$ Department of Physics and Astrophysics, Vrije Universiteit, \\
     De Boelelaan 1081, NL-1081 HV Amsterdam, \\
     The Netherlands\\
$^b$ Department of Physics, \\
North Carolina State University,\\
Raleigh, NC 27695-8202}
\begin{abstract}
\noindent
In the calculation of the anomalous magnetic moment of $W^\pm$ bosons,
we discuss vector anomalies occuring in the fermion loop that spoil the
predictive power of the theory. While the previous analyses were
limited to using essentially the manifestly covariant dimensional
regularization method, we extend the analysis using both the manifestly
covariant formulation and the light-front hamiltonian formulation with
several different regularization methods.  In the light-front dynamics (LFD),
we find that the zero-mode contribution to the helicity zero-to-zero
amplitude for the $W^\pm$ gauge bosons is crucial for the correct 
calculations. Further, we
confirm that the anomaly-free condition found in the analysis of the
axial anomaly can also get rid of the vector anomaly in LFD as well as 
in the manifestly covariant calculations. Our findings in this work may
provide a bottom-up fitness test not only to the LFD calculations but 
also to the theory itself, whether it is any
extension of the Standard Model or an effective field
theoretic model for composite systems. 
\end{abstract}

\date{\today}

\maketitle

\section{Introduction}
\label{sect.I}
\noindent
Anomalies betray the true quantal character of a quantized field
theory. Because they are invariably associated with divergent amplitudes,
their evaluation has proven to be complicated, at times even leading to
enigmatic results~\cite{SV67}.
Nowadays there exists a vast literature on the subject and perhaps a
consensus has been reached~\cite{Bert96}. By definition an anomaly is a
radiative correction that violates a symmetry of the classical Lagrangian
and usually involves counting infinities whether it is due to
ultraviolet infinities or an infinite number of degrees of freedom
~\cite{Jac99}.
As this breaking of symmetry may bring quantized theory in agreement with
experiment, or, on the contrary spoil the renormalizability of the
theory, Jackiw~\cite{Jac99} discerns with this distinction in mind two
types of infinities: good infinities and bad infinities.

In this work, we are concerned with the bad infinities which cause the
anomalies that ultimately spoil the predictive power of the theory.
In particular, we revisit the vector anomaly which led to the discussions
of the requirement of adding a contact term to the magnetic moment
~\cite{DKMT1}
and the superconvergence relations~\cite{DKMT2}, {\it etc.},
in an effort to rescue the theory long time ago.
As the overwhelming majority of investigations in
quantum field theory use the Wick rotation to Euclidian space in the
actual computation of amplitudes, the question whether the occurrence
of anomalies may depend on the formulation of the theory in Euclidian
space remains largely unanswered. Also, it has not yet been so
clear how the appearance of vector anomalies changes depending on the
quantization methods as well as their associated regularization methods.
In this paper we begin to give an
answer, which cannot be considered final, as we concentrated on a single
example, the electromagnetic form factors of the $W^\pm$ bosons in the
Standard Model (SM). We do so by adopting light-front dynamics (LFD),
a Hamiltonian form of dynamics formulated in Minkowski space,
which appears to be another promising technique in  
computing physical observables.
The results are compared to a conventional manifestly covariant
calculation.

It is commonly 
believed~\cite{GB,NEL,SCH} that LFD, if treated
carefully, gives
correct results for $S$-matrix elements, in complete agreement with
manifestly covariant perturbative calculations. The main advantage of LFD
is expected to reveal itself in its application to bound-state problems,
but any calculational tool used in such calculations must prove its
correctness in an application, if any exists, to perturbation
theory. Such a fitness test is particularly needed for regularization
methods devised to render physical quantities finite and pave the way for
renormalization of the theory. Besides its advantages for the calculation
of bound states, one may study LFD for its own sake as an approach to
quantum field theory different from the manifestly covariant formulation,
which may shed some new light on old problems. Here we demonstrate that
anomalies may occur in the LF formulation due to 
the fact that the integrals defining the radiative corrections have a
different character in Minkowski space than in Euclidian space. 
We have chosen the case study of the electromagnetic form factors of the 
$W^\pm$ gauge bosons in the first place to remove any
doubt one may have concerning the applicability of the LF gauge to the
Standard Model. 

As this paper contains many details about the different ways of
calculating and regularizing the amplitudes, we first present a
brief summary of our main results in this section and the details 
of the calculations in the following sections.

\vspace{2ex}

\noindent{\em Summary of the main results}\\
The Lorentz-covariant and gauge-invariant CP-even electromagnetic
$\gamma W^+ W^ -$ vertex is defined~\cite{BGL72,CN87} by
\begin{eqnarray}
 \Gamma^\mu_{\alpha\beta} & = & i\,e \left\{ A[ (p+p')^\mu g_{\alpha\beta}
 +2(g^\mu_\alpha q_\beta - g^\mu_\beta q_\alpha)] \right.
\nonumber \\
 & &\left. + (\Delta \kappa)( g^\mu_\alpha q_\beta - g^\mu_\beta q_\alpha)
 + \frac{\Delta Q}{2M^2_W} (p+p')^\mu q_\alpha q_\beta\right\},
\label{eq.II.010a}
\end{eqnarray}
where $p(p')$ is the initial(final) four-momentum of the $W$ gauge
boson and $q=p'-p$.  Here, $\Delta \kappa$ and $\Delta Q$ are the
anomalous magnetic and quadrupole moments, respectively. At tree level,
\begin{equation}
 A = 1, \quad \Delta \kappa = 0, \quad \Delta Q = 0,
\end{equation}
for any $Q^2 = -q^2$ because of the point-like nature of $W^\pm$ gauge
bosons.  Beyond the tree level, however,
\begin{equation}
 A = F_1(Q^2), \quad -\Delta \kappa = F_2(Q^2) + 2 F_1(Q^2), \quad -\Delta Q
 = F_3(Q^2),
\end{equation}
where the electromagnetic form factors $F_1, F_2$ and $F_3$ for the
spin-1 particles are defined by the relation to the current matrix
elements:{\it i.e.}, $\Gamma^\mu_{\alpha\beta} = -i\, e
J^\mu_{\alpha\beta}$ and
\begin{eqnarray}
 J^\mu_{\alpha\beta} & = &
 \left\{ -(p+p')^\mu g_{\alpha\beta} \; F_1(Q^2)
 +(g^\mu_\alpha\,q_\beta - g^{\mu}_\beta\,q_\alpha)\;F_2(Q^2)
 + \frac{q_\alpha q_\beta}{2M^2_W}
 (p+p')^\mu F_3(Q^2) \right\}.
\label{eq.II.050a}
\end{eqnarray}
The physical form factors, charge ($G_C$), magnetic ($G_M$), and
quadrupole ($G_Q$), are also related in a well-known way to the form
factors $F_1, F_2$ and $F_3$~\cite{BJ1,BCJ02}:
\begin{eqnarray}
 G_C &=& (1 + \mbox{\small$\frac{2}{3}$}\eta) F_1 +
 \mbox{\small$\frac{2}{3}$}\eta F_2
 +\mbox{\small$\frac{2}{3}$} \eta (1 + \eta) F_3 \nonumber \\
 G_M &=& -F_2 \nonumber \\
 G_Q &=& F_1 + F_2 + (1 + \eta) F_3,
 \label{eq.II.070}
\end{eqnarray}
where $\eta = Q^2/(4M^2_W)$.
Of course, one should note that the charge conjugation symmetry
(or Furry's theorem)~\cite{Furry} does not allow the 
existence of a nonvanishing
vertex of a single
photon with any pair of identical spin-1 neutral particles whether
the neutral particle is a gauge boson such as $\gamma$ and
$Z^0$ or a composite particle such as $\rho^0$, {\it etc.}.

The one-loop contributions to $F_1, F_2$, and $F_3$ have been computed
in the SM over the last thirty years~\cite{BGL72,CN87,Arg+93,PP93}.
Among the one-loop contributions, the fermion-triangle-loop is in
particular singled out because of the anomaly. Due to the unique
coupling factors of this triangle loop, it cannot interfere with any
other loop corrections, whether they are from boson-loops or any other
fermion-loop such as the vacuum-polarization of the photon.  The
absence of higher-order corrections to the triangle anomaly has also
been discussed extensively and is known as the non-renomalization
theorem\cite{non-renormalization}.  Thus, we focus on the
fermion-triangle-loop contribution to the CP-even spin-1 form factors
($F_1, F_2$ and $F_3$) and discuss only the vector anomaly occurring in
this triangle loop.

The technical points involved in the calculations are the interchange of
integrations, shifts of the integration variable in momentum-space
integrals, and the occurrence of surface terms. As these points are
correlated with the method adopted to regularize the amplitudes, we
compare the results given by different regularization schemes.
We consider besides the regularization method used mostly in manifestly
covariant field theory, dimensional regularization (DR${}_4$)~\cite{tHV72},
two other methods, Pauli-Villars regularization (PVR)~\cite{PV49} and
smearing (SMR), a method introduced before~\cite{BCJ01} in the context of
a LF calculation of the form factors of vector mesons.
Although it was demonstrated in Ref.~\cite{BCJ01} that for
any finite value of the regulator mass $\Lambda$ in the SMR treatment
the LFD and manifestly covariant calculations of the form factors fully
agreed, the limit $\Lambda\to \infty$ was not studied and therefore one 
might wonder if the agreement would still hold in that limit. If the
PVR procedure is applied to the struck/spectator fermion we call it
PV${}_1$/PV${}_2$, respectively. 
PVR and SMR can be used in the manifestly covariant approach as well as
in LFD. In the latter approach we introduce dimensional regularization of
the integrals over the transverse momenta (DR${}_2$), which can be used
in Minkowski metric, while DR${}_4$ is restricted to Euclidian integrals.
We denote the helicity matrix elements of the current
by $G^\mu_{h'h} = \langle p',h' | J^\mu | p,h \rangle$. In the manifestly
covariant formulation, expressions for the individual form factors can be
found by inspection of the structure of the matrix elements.  In LFD,
however, we need the explicit relations between matrix elements and form
factors to extract the form factors from the helicity amplitudes.

In the manifestly covariant calculations, the anomalous quadrupole moment
$\Delta Q$ (or $F_3(Q^2)$) is found to
be completely independent from the regularization methods as it must
be, {\it i.e.},
\begin{equation}
 \left( F_3 \right)_{\rm SMR} =
 \left( F_3 \right)_{{\rm PV}_1} =
 \left( F_3 \right)_{{\rm PV}_2} =
 \left( F_3 \right)_{{\rm DR}_4}.
\end{equation}
However, we find that the anomalous magnetic moment $\Delta \kappa$ (or
$F_2(Q^2) + 2F_1(Q^2)$) differs by some fermion-mass-independent
constants depending on the regularization methods:
\begin{eqnarray}
 \left( F_2 + 2F_1 \right)_{\rm SMR} &=&
 \left( F_2 + 2F_1 \right)_{{\rm DR}_4} +
 \frac{g^2 Q_f}{4\pi^2}\left(\frac{1}{6}\right),
\nonumber \\
 \left( F_2 + 2F_1 \right)_{{\rm PV}_1} &=&
 \left( F_2 + 2F_1 \right)_{{\rm DR}_4} +
 \frac{g^2 Q_f}{4\pi^2}\left(\frac{2}{3}\right),
\nonumber \\
 \left( F_2 + 2F_1 \right)_{{\rm PV}_2} &=&
 \left( F_2 + 2F_1 \right)_{{\rm DR}_4} +
 \frac{g^2 Q_f}{4\pi^2}\left(-\frac{1}{3}\right).
\label{cov-result}
\end{eqnarray}
These fermion-mass-independent differences are the {\em vector anomalies}
that we point out.  Unless they are completely cancelled, they would
make a unique prediction of $\Delta \kappa$ impossible.  Within the SM,
they are completely cancelled due to the zero-sum of the charge factor
($\sum_f Q_f =0$) in each generation.

In LFD, we compute the form factors using the following relation
in the $q^+ = q^0 + q^3 =0$ frame,
\begin{eqnarray}
G^+_{++}&=& 2p^+(F_1 + \eta F_3),\;
G^+_{+0}= p^+\sqrt{2\eta} (2F_1 + F_2 + 2\eta F_3),\nonumber\\
G^+_{+-}&=&-2p^+\eta F_3,\;\;
G^+_{00}= 2p^+(F_1 - 2\eta F_2 - 2\eta^2 F_3).
\label{eq.II.100a}
\end{eqnarray}
$G^+_{+-}$ depends on $F_3$ only and $G^+_{++}$ involves only
$F_1$ and $F_3$.  Therefore, the simplest procedure is to solve first
for $F_3$ from $G^+_{+-}$.  Next, $F_1$ is obtained from $G^+_{++}$ and
$F_3$. Finally, $F_2$ can be obtained from the other matrix elements. The
two relevant choices are to use  either $G^+_{+0}$ or $G^+_{00}$ and
consequently we may define
\begin{eqnarray}
(F_2 + 2F_1)^{+0}&=& \frac{1}{p^+}\left[\frac{G^+_{+0}}{\sqrt{2\eta}} +
G^+_{+-} \right],\nonumber \\
(F_2 + 2F_1)^{00}&=& \frac{1}{4p^+\eta}\left[(1+2\eta)G^+_{++} -
G^+_{00}+(1+4\eta)G^+_{+-}\right].
\end{eqnarray}

\noindent
Splitting the covariant fermion propagator into the LF-propagating part
and the LF-instantaneous part, the divergences can show up both in the
valence amplitude containing only the LF-propagating fermions and in
the non-valence amplitude containing a LF-instantaneous fermion. 
Calling the non-zero contribution from the non-valence
part in the $q^+ = 0$ frame the zero-mode, we find that only the
helicity zero-to-zero amplitude $G^+_{00}$ receives a zero-mode
contribution given by
\begin{equation}
\left( G^+_{00} \right)_{\rm z.m.}
= \frac{g^2 Q_f p^+}{2\pi^3 M^2_W} \int^1_0 dx \int d^2 {\vec k}_\perp
\frac{{\vec k}^2_\perp + m^2_1 -x(1-x)Q^2}{{\vec k}^2_\perp + m^2_1 +x(1-x)Q^2}
\neq 0.
\end{equation}
The zero-mode contribution to $G^+_{00}$ is crucial because the
unwelcome divergences from the valence part due to the terms with 
the power $(k_\perp^2)^2$ of the transverse momentum are precisely
cancelled by the same terms with the opposite sign from the zero-mode
contribution. The essential results directly related to the vector
anomaly in DR${}_2$ are summarized as follows:
%
%
\begin{eqnarray}
 (F_2 + 2F_1)^{+0}_{{\rm DR}_2} & = &  (F_2 + 2F_1)_{{\rm DR}_4}
 + \frac{g^2 Q_f}{4\pi^2} \left( \frac{1}{6} \right),
\nonumber \\
 (F_2 + 2F_1)^{00}_{{\rm DR}_2} & = & (F_2 + 2F_1)_{{\rm DR}_4}
 -\frac{g^2 Q_f}{4\pi^2} \left(\frac{1}{2\eta}\right) \left(\frac{1}{3} +
 \frac{2\eta}{9}\right).
\label{LFanomaly}
\end{eqnarray}
The fact that $(F_2 + 2F_1)^{+0}_{{\rm DR}_2}$ and 
$(F_2 + 2F_1)^{00}_{{\rm DR}_2}$ disagree indicates that the vector
anomaly in DR${}_2$ appears as a violation of the rotation symmetry or the
angular momentum conservation ({\it i.e.} the angular condition~\cite{CJ}).

Besides DR${}_2$, we have also applied other regularization methods in
LFD, such as PV${}_1$, PV${}_2$ and SMR, which carry an explicit cutoff
parameter $\Lambda$. Interestingly, in each of these regularization
methods, we find that not only $(F_2 +2F_1)^{+0} = (F_2 + 2F_1)^{00}$ but
also the LF result completely agrees with the corresponding manifestly
covariant result: viz
\begin{equation}
(F_2 +2F_1)^{+0}_{\rm PV1} = (F_2 + 2F_1)^{00}_{\rm PV1} = (F_2
+2F_1)^{cov}_{{\rm PV}_1} 
\label{equivalence}
\end{equation}
where $(F_2 +2F_1)^{cov}_{\rm PV1}$ is the result shown in
Eq.~(\ref{cov-result}).  This proves that the rotation symmetry is
not violated in the regularization methods with an explicit cutoff
$\Lambda$ unlike the above DR${}_2$ case.  However, we note that the
zero-mode contribution in $(F_2 + 2F_1)^{00}$ is crucial to get
an equivalence as shown in
Eq.~(\ref{equivalence}). The details of our calculations including the
interesting consequence in the PV${}_2$ case where the zero-mode is
artificially removed will be presented in this work. 

\vspace{2ex}

The paper is organized as follows.  In Sect.~\ref{sect.II} we briefly
discuss the subtle points concerning the triangle diagrams associated
with the electromagnetic form factors of the $W^\pm$ bosons. We
concentrate on matters pertinent to the particular case at hand and
leave the broad context to the comprehensive review by
Bertlmann~\cite{Bert96}.
Sect.~\ref{sect.III} contains the statement of the problem and its
manifestly covariant formulation and defines our notations and
conventions. Next the results using dimensional regularization applied to
Wick rotated amplitudes (DR${}_4$) along with the results using other
regularizations (SMR, PV${}_1$, PV${}_2$) are given.  In
Sect.~\ref{sect.IV} the LF treatment is presented. Our discussion and
conclusion are presented in Sect.~\ref{sect.VI}. Mathematical details 
for the dimensional regularizations (DR${}_4$, DR${}_2$) are summarized
in Appendix A.

\section{Form factors of $W^\pm$}
\label{sect.II}
\noindent
In a classical paper, Bardeen {\em et al.}~\cite{BGL72} calculated for the
first time the static properties of the $W^\pm$ bosons. Dimensional
regularization is used despite the reservations of the authors concerning
its application in cases where $\gamma_5$ is involved. 
They define the most general CP invariant $\gamma W^+ W^-$ vertex as  
given by Eq.~(\ref{eq.II.010a}) and identify the anomalous magnetic dipole
moment $\Delta\kappa = \kappa -1$ and the anomalous quadrupole moment 
$\Delta Q$.
The corrections from the triangle diagram containing only 
the sum of electron and muon loops in the massless limit 
at $q^2=0$ are given by~\cite{BGL72}
\begin{equation}
 \kappa-1  = -\frac{G_F M^2_W}{2\pi^2\surd 2} \frac{1}{3},\quad \Delta Q  =
 \frac{G_F M^2_W}{2\pi^2\surd 2} \frac{4}{9}.
\label{eq.II.030} 
\end{equation} 
This result sets the standard for the static properties of the weak bosons.
Some years after the publication of this paper, some doubts were raised
concerning the static quantities~\cite{DKMT1,DKMT2}, but its results were
fully
vindicated. Still, some doubts remained, but in Ref.~\cite{BG77} the
gauge invariance of the electro-weak theory regulated using DR${}_4$ was
corroborated and we may consider this the {\em orthodox} point of view.
Still, some textbook authors remain unhappy with the treatment of
matrix elements involving $\gamma_5 $~\cite{IZ80} and prefer Pauli-Villars
regularization for those cases.

Calculations beyond the limit of massless fermions were done
later~\cite{CN87,Arg+93}. We use the notation of these references for
the mass ratios and the integrals given by
\begin{equation}
 \Delta = m^2_1/M^2_W,\quad {\cal E} = m^2_2/M^2_W,\quad
F=1-\Delta+{\cal E}, \quad
 I_n = \int^1_0 dt \frac{t^n}{t^2 - F t + {\cal E}}. 
\label{eq.II.040}
\end{equation}
The mass $m_2$ is the mass of the spectator fermion, while $m_1$ is the 
mass of the fermion involved in the $\gamma f\bar{f}$ vertex of the 
triangle
diagram.
The authors of Refs.~\cite{CN87,Arg+93} again find that the gauge
invariance
is maintained using DR${}_4$ and mention that the fermion contribution
vanishes for the fermion families in the massless limit approximation
owing to the anomaly-free condition in the SM.

As we want to connect the results of the present paper with our results
obtained for the form factors of vector mesons, we define our conventions
as in Ref.~\cite{BCJ02}.
The Lorentz-invariant electromagnetic form factors $F_1$, $F_2$, and $F_3$
for spin-1 bosons are defined by 
Eq.~(\ref{eq.II.050a}).
The orthodox point of view being that DR${}_4$ is a gauge-invariant
regularization, we use the light-front gauge as we are interested in
applying light-front dynamics to the calculation of the current matrix
elements. 

\subsection{Current Matrix Elements in the Light-Front Gauge}
\label{sect.II.01}
\noindent
The relation between the
covariant form factors $F_i$ and the current matrix elements 
$G^\mu_{h' h} = \la p',h'|J^\mu|p,h\ra$ is given by
\begin{eqnarray}
 G^\mu_{h' h} &=& -\ep^{*}_{h'}\cdot\ep_h(p+p')^\mu F_1(Q^2) \nonumber \\
 & & + (\ep^\mu_h\;q\cdot\ep^{*}_{h'} 
  - \ep^{*\mu}_{h'}\;q\cdot\ep_h)F_2(Q^2) \nonumber \\
 & & + \frac{(\ep^{*}_{h'}\cdot q)(\ep_h\cdot q)}{2M^2_W} 
(p+p')^\mu F_3(Q^2).
\label{eq.IV.010}
\end{eqnarray}
Here, the general form of the LF polarization vectors~\cite{BCJ02}
is given by
\begin{equation}
 \left.
 \begin{array}{c}
 \varepsilon_{\rm ff} (p^+,p^1,p^2 ;+) \\
 \varepsilon_{\rm ff} (p^+,p^1,p^2 ;0) \\
 \varepsilon_{\rm ff} (p^+,p^1,p^2 ;-)
 \end{array}
 \right\} = \left\{
 \begin{array}{c}
 \left( 0, \frac{p^r}{p^+}, \frac{-1}{\surd 2}, \frac{-i}{\surd 2},
 \right) \\
 \left( \frac{p^+}{m}, \frac{\vec{p}^2_\perp - m^2}{2 m p^+},
 \frac{p^1}{m}, \frac{p^2}{m}
 \right) \\
 \left( 0, \frac{p^l}{p^+}, \frac{ 1}{\surd 2}, \frac{-i}{\surd 2} \right)
 \end{array} \right. ,
 \label{eq.II.080}
\end{equation}
where $p^r (p^l) = \mp (p_x \pm ip_y)/\surd 2$.
As it is well-known that the plus current $J^+$ suffers the least from
zero-mode problems in LFD~\cite{BCJ01} , we shall only use this current
component here. For the evaluation of the matrix elements,
$G^+_{h' h} = \langle p',h'|J^+|p,h \rangle$,
we need to define the kinematics for the reference frame that we are
going to take.

\vspace{2ex}

\noindent{\em Kinematics}\\
We use the notation 
$p^\mu = (p^+, p^-, p_x, p_y) = (p^+, p^-, \vec{p}_\perp)$ and the metric
$p' \cdot p = (p^{\prime +} p^- + p^{\prime -} p^+)/2 - p'_x p_x - p'_y p_y$.
\noindent
In the literature, usually the reference frames are taken
as the ones where $q^+ = 0\; (q^2=q^+q^-- \vec{q}^{\,2}_\perp<0)$.
A particular useful one  is the frame where $q^+=0, q_x=Q, q_y=0$,
and $\vec{p}_\perp=-\vec{p'}_\perp$:
\begin{eqnarray}
 q^\mu & = & (0,0,Q, 0),\nonumber \\
 p^\mu& =& (M_W\sqrt{1+\eta}, M_W\sqrt{1+\eta}, -Q/2,0),\nonumber \\
 p'^\mu&  = & (M_W\sqrt{1+\eta}, M_W\sqrt{1+\eta}, Q/2,0).
\label{eq.II.090}
\end{eqnarray}
The corresponding polarization vectors are obtained by substituting
these four vectors in Eq.~(\ref{eq.II.080}).

The angular condition for the spin-1 system can be
obtained from the explicit representations of the helicity amplitudes
in terms of the physical form factors.  
Using Eq.~(\ref{eq.IV.010}), one can obtain Eq.~(\ref{eq.II.100a}) in the
kinematics we have chosen.
In the limit $Q^2\to 0$ one can retrieve the static quantities in the
following way
\begin{equation}
 F_1(0) = \frac{G^+_{++}}{2 p^+} = \frac{G^+_{00}}{2 p^+},\; 
 F_2(0) + 2F_1(0) =  \frac{G^+_{+0}}{p^+ \sqrt{2\eta}},\;
 F_3(0) = -\frac{G^+_{+-}}{2 p^+\eta},
\label{eq.II.110}
\end{equation}
where the appropriate limit $\eta\to 0$ must be taken. For arbitrary
values of $Q^2$, we obtain by inverting the relations~(\ref{eq.II.100a})
\begin{eqnarray}
 F_1 & = & \frac{1}{2 p^+} \;\left[ G^+_{++} + G^+_{+-} \right],
\nonumber \\
 F^{+0}_2 & = & \frac{1}{p^+}
 \;\left[-G^+_{++} + \frac{1}{\sqrt{2\eta}}G^+_{+0} \right],
\nonumber \\
 F^{00}_2 & = & \frac{1}{4\eta p^+} \;
 \left[ (1-2\eta)G^+_{++} + G^+_{+-} -G^+_{00}\right],
\nonumber \\
 F_3 & = & - \frac{1}{2p^+\eta} \; G^+_{+-},
\label{eq.II.120}
\end{eqnarray}
where the upper indices on $F_2$ indicate which of the two matrix
elements,
$G^+_{+0}$ or $G^+_{00}$, is used to determine it.  The angular condition
relating the four helicity amplitudes is $F^{+0}_2 =  F^{00}_2$ and was
given before~\cite{BJ02} in terms of the current matrix elements
\begin{equation}
 (1+2\eta) G^+_{++} + G^+_{+-} -\sqrt{8\eta}G^+_{+0} - G^+_{00}=0.
\label{eq.II.130}
\end{equation}
\noindent
In this paper, we are concerned with the magnetic dipole and the electric
quadrupole form factors of the $W$ bosons. If the angular condition
is violated, they will not be determined unambiguously by the current
matrix elements either.

\section{Manifestly covariant calculation}
\label{sect.III}
\begin{figure}
\centerline{\epsfig{figure=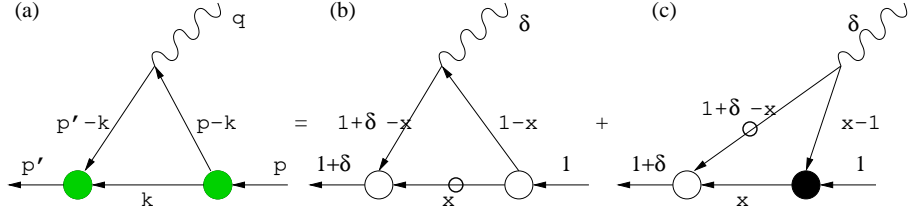,width=120mm}}
\caption{The covariant triangle diagram (a) is represented as the sum
of a LF valence diagram (b) defined in the region $0<k^+<p^+$ and the 
nonvalence diagram (b) defined in $p^+<k^+<p'^+$.
$\delta=q^+/p^+=p'^+/p^+-1$. The white and black blobs at
the boson-fermion vertices in (b) and (c) represent the 
LF wave-function and non-wave-function vertices, respectively,
for a bound-state boson. In the SM, however, $W^\pm$ gauge bosons
are point particles and those blobs reduce to point vertices.
The small circles in (b) and (c) represent the (on-shell)
mass pole of the fermion propagator determined from the
$k^-$-integration. 
\label{Fig1}}
\end{figure}

\noindent
The current matrix element $G^\mu_{h'h}$ of a 
spin-1 particle with constituents with masses $m_1$ and $m_2$ is
obtained from the
covariant diagram, Fig.~\ref{Fig1}(a), and given by
\begin{equation}
 G^\mu_{h'h} = -i g^2 Q_f \int\frac{d^4k}{(2\pi)^4} \frac{T^\mu_{h'h}}
 {[(k-p)^2-m^2_1+i\vep][k^2-m^2_2+i\vep][(k-p')^2-m^2_1+i\vep]},
\label{eq.III.010}
\end{equation}
where $T^\mu_{h'h}$ is the trace term of the fermion propagators and
$g^2 = G_F M^2_W / \sqrt{2} $ in the SM.  The charge factor $Q_f$
includes the color factor $N_c$ if the fermion loop is due to a quark.

To regularize the covariant fermion triangle-loop in ($3+1$) dimension,
different choices can be made. One may use dimensional regularization
(DR)~\cite{tHV72} or use the classical Pauli-Villars
method~\cite{PV49}. We consider another method too, so-called smearing
(SMR),
where we replace the point photon-vertex $\gamma^\mu$ by a non-local
photon-vertex $S_\Lambda(p)\gamma^\mu S_\Lambda(p')$, where
$S_\Lambda(p)=\Lambda^2/(p^2-\Lambda^2+i\vep)$ and $\Lambda$ plays the
role of a momentum cut-off similar but not identical to Pauli-Villars
regularization.

\subsection{Details of the Computation}
\label{sect.III.1}
\noindent
We first determine the trace in the covariant expression. It is
\begin{eqnarray}
 T^\mu_{h'h} & = &
 {\rm Tr}
[\not\!\ep^{*}_{h'}(1-\gamma_5)(\not\! k + m_2)\not\!\ep_{h}(1-\gamma_5)
(\not\! k-\not\!p+m_1)\gamma^\mu(\not\!k -\not\!p'+ m_1)]
\nonumber\\
 & = &  2 (T^\mu_{h'h})_{CP^+} + 2 (T^\mu_{h'h})_{CP^-} \nonumber \\
 & = &  \;
 2 {\rm Tr}
[\not\!\ep^{*}_{h'}\not\! k\not\!\ep_{h}
(\not\! k-\not\!p+m_1)\gamma^\mu(\not\! k-\not\!p'+ m_1)],
\nonumber\\
 &  &  +
 2 {\rm Tr}
[\gamma_5 \not\!\ep^{*}_{h'}\not\!k\not\!\ep_{h}
(\not\! k-\not\!p+m_1)\gamma^\mu(\not\!k -\not\!p'+ m_1)].
 \label{eq.III.020}
\end{eqnarray}
The two traces determine two parts of the triangle diagram: a $CP$-even
one and a $CP$-odd one. We shall not pursue the latter here.
The $CP$-even part is found to be
\begin{eqnarray}
(T^\mu_{h'h})_{CP^+} & = &  4 \left\{
 \, k^\mu \left[ 4\, \ep^{*}_{h'} \cdot k \, \ep_{h}\cdot k
 -  \ep^{*}_{h'}\cdot p\,  \ep_h\cdot p' 
 -  \ep^{*}_{h'}\cdot\ep_h\, (k^2 - p\cdot p' + m^2_1)\right]\right.
\nonumber\\
 & & \left. + \, p^\mu \left[
  - 2  \ep^{*}_{h'}\cdot k\, \ep_h\cdot k 
 +  \ep^{*}_{h'}\cdot k\, \ep_h\cdot p' 
 +  \ep^{*}_{h'}\cdot\ep_h\, (k^2 - k\cdot p')\right]\right.
\nonumber\\
 & &\left.  + \, p^{\prime\mu} \left[
  - 2  \ep^{*}_{h'}\cdot k\, \ep_h\cdot k 
 +  \ep^{*}_{h'}\cdot p\, \ep_h\cdot k 
 +  \ep^{*}_{h'}\cdot\ep_h\, (k^2 - k\cdot p)\right]\right.
\nonumber\\
 & &\left.  - \,\ep^\mu_{h}\left[
 \ep^{*}_{h'}\cdot k\, (k^2 - 2 k\cdot p + p'\cdot p - m^2_1)
 + \ep^{*}_{h'}\cdot p\, (k^2 - k\cdot p') \right]\right.
\nonumber\\
 & & \left. - \,\ep^{* \mu}_{h'} \left[
 \ep_{h}\cdot k\, (k^2 - 2 k\cdot p' + p'\cdot p - m^2_1)
 + \ep \cdot p' \, (k^2 - k\cdot p) \right] \right\}.
\nonumber \\
\label{eq.III.030}
\end{eqnarray}
\noindent 
We use the Feynman parametrization for the three propagators, {\it e.g.},
\begin{eqnarray}
 \frac{1}{D_k D_0 D'_0}&=& 2\int^1_0 dx\int^{1-x}_0 dy
 \frac{1}{[x D_0 + y D'_0  + (1-x-y) D_k]^3}.
\label{eq.III.040}
\end{eqnarray}
We shall compute the integrals over the momentum by shifting the
integration variable from $k$ to $k' = k - (xp + yp')$, although we are
aware of the fact that a shift is not permitted unless the integral can
be shown to diverge at most logarithmically. We shall return to this
point later. Then one finds a denominator that is an even function of $k'$:
\begin{equation}
 D  =  [k - (x p + y p')]^2 + (x+y)(1-x-y) M^2_W  -(x+y) m^2_1 - (1-x-y) m^2_2
 + xyq^2.
\label{eq.III.050}
\end{equation}
We thus can write the denominator function in the form
\begin{equation}
 D = k^{\prime 2} - M^2_W C^2_{\rm cov}.
\label{eq.III.060}
\end{equation}
Using the notation given in Eq.~(\ref{eq.II.040}), the mass function is
\begin{equation}
 - C^2_{\rm cov} = (x+y)(1-x-y) -(x+y)\Delta - (1-x-y){\cal E} - 4 
xy\eta.
\label{eq.III.070}
\end{equation}

The shift in momentum variable must be applied to the trace too. Doing
so and discarding the terms odd in $k'$ gives the new numerator
given by
\begin{eqnarray}
S^\mu_{h'h} := (T^\mu_{h'h})_{CP^+} & = & 
 4\left\{\, \ephp\cdot\eph\,
 \left[
  p^\mu \left(\frac{1-x}{2} k^{\prime 2} +
  \{(1-x)(x+y)^2 - y\} M^2_W - (1-x)xyq^2 - xm_1^2\right)
 \right. \right.
 \nonumber \\
 & & \left. \left. \quad\quad\quad +
 p^{\prime \mu} \left(\frac{1-y}{2} k^{\prime 2} +
  \{(1-y)(x+y)^2 - x\} M^2_W - (1-y)xyq^2 - ym_1^2\right)
 \right]  \right.
\nonumber\\
 & & \left. + \ephp \cdot p \, \eph^\mu
 \left[ -\frac{1+x}{2} k^{\prime 2} - (x+y-1) (x^2 +(1+x) y)M^2_W
 +x^2 yq^2 + x m^2_1
 \right] \right.
\nonumber\\
 & &\left.  + \eph \cdot p'\,   \ep_{h'}^{*\mu}
 \left[ -\frac{1+y}{2} k^{\prime 2} - (x+y-1) (y^2 +(1+y) x)M^2_W
 +x y^2 q^2 + y m^2_1
 \right] \right.
\nonumber\\
 & &\left.  +  \ephp \cdot p\;  \eph \cdot p'\; (2xy)
 \left[ (2x-1) p^\mu + (2y-1) p^{\prime\mu} 
  \right] \rule{0mm}{3ex} \right\} .
\label{eq.III.080}
\end{eqnarray}
From Eq.~(\ref{eq.III.050}), we see that 
part of 
the denominator is symmetric under
interchange of $p$ and $p'$ and $x$ and $y$. 
It is not difficult to show that only the numerator symmetric in $x$ and 
$y$ can survive the integration.
Therefore, we may symmetrize the formal expression~(\ref{eq.III.080}). 
We find for the symmetrized trace
\begin{equation}
 (S^\mu_{h'h})_{\rm symm} = 
  \ephp\cdot\eph\,  (p + p')^\mu f_1
  +(\ephp\cdot p\, \eph^\mu + \eph\cdot p'\, \ep_{h'}^{*\mu}) f_2
 + \ephp\cdot p\;  \eph \cdot p' (p + p')^\mu f_3 ,
\label{eq.III.090}
\end{equation}
where the functions $f_1$, $f_2$, and $f_3$ are
\begin{eqnarray}
  f_1= f^1_1 k^{\prime 2} + f^0_1 & = &  (2-x-y)\; k^{\prime 2} \nonumber \\
 & & - 2 [(x+y)(1-x-y)^2 M^2_W + (2-x-y)xyq^2 + (x+y)m^2_1] ,
\nonumber\\
 f_2 = f^1_2 k^{\prime 2} + f^0_2 & = & -(2+x+y) k^{\prime 2}\nonumber \\
 & &  + 2 (x+y)[(1-(x+y)^2)M^2_W + xyq^2 + m^2_1] ,
\nonumber\\
 f_3 & = & 8 x y (x+y-1) .
\label{eq.III.100}
\end{eqnarray}
Comparing Eq.~(\ref{eq.III.100}) with Eq.~(\ref{eq.IV.010}), we can read
off the integrands for the three form factors.
\noindent
In the manifestly covariant calculation, we first obtain the form factors
$F_i(i=1,2,3)$ using dimensional regularization DR${}_4$:
\begin{eqnarray}
 F_1(Q^2)&=& \frac{g^2 Q_f}{4\pi^2} \int^1_0dx\int^{1-x}_0dy
 \biggl\{
 -(2-x-y) \left[\frac{1}{\ep} - \gamma - \frac{1}{2} +
  \ln \frac{4\pi\mu^2}{M^2_W C^2_{\rm cov}}\right]
 + \frac{1}{2} \frac{f^0_1}{M^2_W C^2_{\rm cov}}
 \biggr\}
 \nonumber\\
 F_2(Q^2)&=& \frac{g^2 Q_f}{4\pi^2} \int^1_0dx\int^{1-x}_0dy
 \biggl\{
 (2+x+y) \left[\frac{1}{\ep} - \gamma - \frac{1}{2} +
  \ln \frac{4\pi\mu^2}{M^2_W C^2_{\rm cov}}\right]
 + \frac{1}{2} \frac{f^0_2}{M^2_W C^2_{\rm cov}}
 \biggr\}
\nonumber\\
 F_3(Q^2)&=& \frac{g^2 Q_f}{4\pi^2} \int^1_0dx\int^{1-x}_0dy\;
 \frac{8xy(x+y-1)}{C^2_{\rm cov}},
\label{eq.III.120}
\end{eqnarray}
where $\ep$ is defined in the Appendix.

In order to obtain the form factors using PVR or SMR, one introduces
fictitious particles with mass $\Lambda$.  In true Pauli-Villars
regularization one replaces the amplitude $M^\mu(m_1,m_2)$ by the
amplitude $M^{\rm Reg} = M^\mu(m_1,m_2) - M^\mu(\Lambda_1, m_2)$
or $M^{\rm Reg} = M^\mu(m_1,m_2) - M^\mu(m_1, \Lambda_2)$, depending
on whether one chooses to replace the propagator of the struck fermion
or the spectator fermion, respectively, by the combination 
$(\not\!p - m)^{-1} - (\not\!p - \Lambda)^{-1}$. 
The SMR procedure~\cite{BCJ01}
consists in replacing the $\gamma f\bar{f}$-vertex $\gamma^\mu$ by
the vertex $S_\Lambda (p') \gamma^\mu S_\Lambda (p)$ using the SMR
function $S_\Lambda (p) = \Lambda^2 / (p^2 - \Lambda^2 + i\varepsilon)$.
For the purpose of making the situation transparent, we define the function
\begin{equation}
 -C^2(m_a,m_b) = (x+y)(1-x-y)M^2_W -xm^2_a -ym^2_b - (1-x-y)m^2_2 +
xy\;q^2,
\label{eq.III.130}
\end{equation}
where $C^2(m_1,m_1) = M^2_W C^2_{\rm cov}$.
The full result using SMR for the form factors of a vector
meson (or spin-1 bound-states) can be found in Ref.~\cite{BCJ02}.
The SMR result amounts to the following replacements of the
logarithmic term and the factor $1/C^2_{\rm cov}$ in the nonlogarithmic
parts
in Eq.~(\ref{eq.III.120}):
\begin{eqnarray}
 \ln(C^2_{\rm cov}) & \to &
 \ln\left(
 \frac{C^2(m_1,m_1) C^2(\Lambda,\Lambda)}{C^2(m_1,\Lambda) C^2(\Lambda,m_1)}
 \right),
\nonumber \\
 \frac{1}{M^2_W C^2_{\rm cov}} & \to &
 \frac{1}{C^2(m_1,m_1)}-\frac{1}{C^2(m_1,\Lambda)}-
 \frac{1}{C^2(\Lambda, m_1)} + \frac{1}{C^2(\Lambda,\Lambda)}.
\label{eq.III.140}
\end{eqnarray}
Note that the substitution
of $\Lambda$ for $m_1$ only affects the denominators. In the case of
PVR, the propagator mass is replaced, affecting
both the numerator and the denominator. The result is given by the
substitutions in the integrands 
${\cal F}_i$
defining the form factors $F_i$
\begin{equation}
 {\cal F}_i (m_1,m_2) \to {\cal F}_i (m_1,m_2)  - {\cal F}_i (\Lambda_1,m_2) , 
\quad (i=1,2,3)
\label{eq.III.150}
\end{equation}
for Pauli-Villars in the fermion line connected to the photon and
\begin{eqnarray}
 {\cal F}_i (m_1,m_2) \to {\cal F}_i (m_1,m_2)  - {\cal F}_i
(m_1,\Lambda_2) , \quad (i=1,2,3)
\label{eq.III.160}
\end{eqnarray}
for the regularization involving the spectator only. 
The infinities that plague $F_1$ and $F_2$, represented by the
terms in 
Eq.~(\ref{eq.III.120}) 
containing the factor $1/\epsilon$ are 
also found in SMR or PVR if the limit
$\Lambda\to\infty$ is taken. In particular, we find for the SMR case
\begin{equation}
 - \ln\left(
 \frac{C^2(m_1,\Lambda) C^2(\Lambda,m_1)}{C^2(m_1,m_1) C^2(\Lambda,\Lambda)}
 \right)
 \to
 - \ln\left( \frac{\Lambda^2}{M^2_W}\right) +
 \ln\left(\frac{x+y}{xy}\right) + \ln(C^2_{\rm cov}),
\label{eq.III.170}
\end{equation}
while the nonlogarithmic terms containing $\Lambda$ vanish. 

For the two PVR cases, $m_1 \to \Lambda_1$ (PV${}_1$) and $m_2 \to
\Lambda_2$ (PV${}_2$), we find the logarithm to be replaced by
\begin{eqnarray}
  \ln(C^2_{\rm cov}) & \to & - \ln\left(\frac{\Lambda^2_1}{M^2_W}\right)
 -\ln(x+y) +\ln(C^2_{\rm cov}),
\nonumber \\
  \ln(C^2_{\rm cov}) & \to & - \ln\left(\frac{\Lambda^2_2}{M^2_W}\right)
 -\ln(1-x-y) +\ln(C^2_{\rm cov}),
\label{eq.III.180}
\end{eqnarray}
respectively. The nonlogarithmic term 
in the integrand of $F_1$($F_2$) gets corrected by $-1$($+1$) for 
PV${}_1$. For PV${}_2$, the corrections to 
the nonlogarithmic terms are zero in the limit $\Lambda_2\to\infty$ 
because $m_2$ does not occur in the numerators.

Clearly, the infinity in the DR case is recovered as a term $\ln \Lambda$
in SMR or PVR. There appear, however, finite terms
that contribute to the finite parts of the form factors $F_1$ and $F_2$.
These terms are independent of the masses of the fermions and, when integrated
over $x$ and $y$, appear as pure numbers.

\subsection{Summary of Covariant Results}
\label{sect.III.2}

\noindent
{\em\bf DR${}_4$}\\
\noindent
The three form factors obtained in DR${}_4$ can be summarized as
\begin{eqnarray}
 F_1(Q^2)&=& \frac{g^2 Q_f}{4\pi^2}
 \biggl\{
 -\frac{2}{3} \left(\frac{1}{\ep} - \gamma - \frac{1}{2} \right) +
 \int^1_0dx\int^{1-x}_0dy
 \left[
 -(2-x-y) \ln \frac{4\pi\mu^2}{M^2_W C^2_{\rm cov}}
 + \frac{1}{2} \frac{f^0_1}{M^2_W C^2_{\rm cov}}\right]
 \biggr\}
 \nonumber\\
 F_2(Q^2)&=& \frac{g^2 Q_f}{4\pi^2}
 \biggl\{
 \frac{4}{3} \left(\frac{1}{\ep} - \gamma - \frac{1}{2} \right) +
 \int^1_0dx\int^{1-x}_0dy
 \left[
  (2+x+y) \ln \frac{4\pi\mu^2}{M^2_W C^2_{\rm cov}}
 + \frac{1}{2} \frac{f^0_2}{M^2_W C^2_{\rm cov}}\right]
 \biggr\}
\nonumber\\
 F_3(Q^2)&=&  \frac{g^2 Q_f}{4\pi^2} \int^1_0dx\int^{1-x}_0dy\;
 \frac{8xy(x+y-1)}{C^2_{\rm cov}}.
\end{eqnarray}
The physical quantity, -$\Delta\kappa$, corresponds to
\begin{eqnarray}
 2 F_1(Q^2) + F_2(Q^2) &=& \frac{g^2 Q_f}{4\pi^2}
 \int^1_0dx\int^{1-x}_0dy
 \left[
 -(2-3x-3y) \ln \frac{4\pi\mu^2}{M^2_W C^2_{\rm cov}}
 + \frac{1}{2} \frac{2f^0_1 + f^0_2}{M^2_W C^2_{\rm cov}}\right]
\nonumber\\
&=& \frac{g^2 Q_f}{4\pi^2}
 \int^1_0dx\int^{1-x}_0dy
 \left[
 (2-3x-3y) \ln C^2_{\rm cov}
 + \frac{1}{2} \frac{2f^0_1 + f^0_2}{M^2_W C^2_{\rm cov}}\right].
\label{eq.III.201}
\end{eqnarray}
The singular part of $2F_1(Q^2) + F_2(Q^2)$ is cancelled out 
and the dependence on the mass scale $\mu$ also vanishes
when integrated over $x$ and $y$. 
Changing the variables $x$ and $y$ to $t=x+y$ and $u=x-y$,
the relevant integral becomes
\begin{equation}
 \frac{1}{2}\int^1_0 dt \int^t_{-t} du (2-3t) \ln \frac{4\pi\mu^2}{M^2_W}
= \int^1_0 dt \; t(2-3t) \ln \frac{4\pi\mu^2}{M^2_W}
\label{eq.III.230}
\end{equation}
which vanishes for any mass scale $\mu$.
This indicates that $2F_1 + F_2$
is defined by a {\em conditionally convergent} integral.
At $q^2 = 0$, 
the part containing $\ln(C^2_{\rm cov})$ can be cast in the form of a
combination of integrals $I_n$ by performing an integration by parts;
\begin{eqnarray}
F_2(0) + 2 F_1(0) & = &
-\frac{g^2 Q_f}{4\pi^2} \int^1_0 dt \left\{ (3t^2 -2t) \ln 
(t^2-tF+{\cal E})
+  \frac{3t^4-4t^3+t^2(1+\Delta)}{t^2-tF+{\cal E}} \right\} 
\nonumber \\
& = & -\frac{g^2 Q_f}{4\pi^2} \int^1_0 dt \frac{t^4-(F-2)t^3+
(2\Delta-{\cal E})t^2}{t^2-tF+{\cal E}}.
\label{eq.III.added}
\end{eqnarray}
In this way, we recover the well known results for the physical quantities 
~\cite{CN87}
\begin{eqnarray}
 F_2(0) + 2 F_1(0) & = &  
 -\frac{g^2 Q_f}{4 \pi^2} [I_4 - (F-2) I_3 + (2\Delta - {\cal E}) 
I_2],
\nonumber \\
 F_3(0) & = &  
 -\frac{g^2 Q_f}{4 \pi^2} \;\frac{4}{3} [I_3 - I_4].
\label{eq.III.220}
\end{eqnarray}

Since the form factor $F_3$ needs no regularization, we do not include it 
in the
summary below, where we give the results for the other regularizations.
\vspace{3ex}

\noindent
{\em\bf SMR}\\
\begin{eqnarray}
 F_1(Q^2)&=& \frac{g^2 Q_f}{4\pi^2}
 \biggl\{
 -\frac{2}{3} \ln \frac{\Lambda^2}{M^2_W} + \frac{31}{18} +
 \int^1_0dx\int^{1-x}_0dy
 \left[
 -(2-x-y) \ln \frac{1}{C^2_{\rm cov}}
 + \frac{1}{2} \frac{f^0_1}{M^2_W C^2_{\rm cov}}\right]
 \biggr\}
 \nonumber\\
 F_2(Q^2)&=& \frac{g^2 Q_f}{4\pi^2}
 \biggl\{
 \frac{4}{3} \ln \frac{\Lambda^2}{M^2_W} - \frac{59}{18} +
 \int^1_0dx\int^{1-x}_0dy
 \left[
  (2+x+y) \ln \frac{1}{ C^2_{\rm cov}}
 + \frac{1}{2} \frac{f^0_2}{M^2_W C^2_{\rm cov}}\right]
 \biggr\}
\nonumber \\
 2 F_1(Q^2) + F_2(Q^2) & = & \frac{g^2 Q_f}{4\pi^2}
 \biggl\{
  \frac{1}{6} +
 \int^1_0dx\int^{1-x}_0dy
 \left[
 (2-3x-3y) \ln C^2_{\rm cov}
 + \frac{1}{2} \frac{2f^0_1 + f^0_2}{M^2_W C^2_{\rm cov}}\right]
 \biggr\},
\label{eq.III.240}
\end{eqnarray}

\noindent
{\em\bf PV${}_1$}\\
\begin{eqnarray}
 F_1(Q^2)&=& \frac{g^2 Q_f}{4\pi^2}
 \biggl\{
 -\frac{2}{3} \ln \frac{\Lambda^2}{M^2_W} + \frac{8}{9} +
 \int^1_0dx\int^{1-x}_0dy
 \left[
 -(2-x-y) \ln \frac{1}{C^2_{\rm cov}}
 + \frac{1}{2} \frac{f^0_1}{M^2_W C^2_{\rm cov}}\right]
 \biggr\}
 \nonumber\\
 F_2(Q^2)&=& \frac{g^2 Q_f}{4\pi^2}
 \biggl\{
 \frac{4}{3} \ln \frac{\Lambda^2}{M^2_W} - \frac{10}{9} +
 \int^1_0dx\int^{1-x}_0dy
 \left[
 (2+x+y) \ln \frac{1}{C^2_{\rm cov}}
 + \frac{1}{2} \frac{f^0_2}{M^2_W C^2_{\rm cov}}\right]
 \biggr\}
\nonumber \\
 2 F_1(Q^2) + F_2(Q^2) & = & \frac{g^2 Q_f}{4\pi^2}
 \biggl\{
  \frac{2}{3} +
 \int^1_0dx\int^{1-x}_0dy
 \left[
 (2-3x-3y) \ln C^2_{\rm cov}
 + \frac{1}{2} \frac{2f^0_1 + f^0_2}{M^2_W C^2_{\rm cov}}\right]
 \biggr\},
\label{eq.III.250}
\end{eqnarray}

\noindent
{\em\bf PV${}_2$}\\
\begin{eqnarray}
 F_1(Q^2)&=& \frac{g^2 Q_f}{4\pi^2}
 \biggl\{
 -\frac{2}{3} \ln \frac{\Lambda^2}{M^2_W} + \frac{8}{9} +
 \int^1_0dx\int^{1-x}_0dy
 \left[
 -(2-x-y) \ln \frac{1}{C^2_{\rm cov}}
 + \frac{1}{2} \frac{f^0_1}{M^2_W C^2_{\rm cov}}\right]
 \biggr\}
 \nonumber\\
 F_2(Q^2)&=& \frac{g^2 Q_f}{4\pi^2}
 \biggl\{
 \frac{4}{3} \ln \frac{\Lambda^2}{M^2_W} - \frac{19}{9} +
 \int^1_0dx\int^{1-x}_0dy
 \left[
 (2+x+y) \ln \frac{1}{C^2_{\rm cov}}
 + \frac{1}{2} \frac{f^0_2}{M^2_W C^2_{\rm cov}}\right]
 \biggr\}
\nonumber \\
 2 F_1(Q^2) + F_2(Q^2) & = & \frac{g^2 Q_f}{4\pi^2}
 \biggl\{
 - \frac{1}{3} +
 \int^1_0dx\int^{1-x}_0dy
 \left[
 (2-3x-3y) \ln C^2_{\rm cov}
 + \frac{1}{2} \frac{2f^0_1 + f^0_2}{M^2_W C^2_{\rm cov}}\right]
 \biggr\}.
\label{eq.III.260}
\end{eqnarray}

\section{Light-front calculation}
\label{sect.IV}
\noindent
The covariant amplitude shown in Fig.~\ref{Fig1}(a) is in general
believed to be equivalent to the sum of the LF valence diagram (b) and the
nonvalence diagram (c), where the notation $\delta=q^+/p^+=p'^+/p^+-1$
is used. For amplitudes that are defined by convergent integrals this
belief is well founded~\cite{NEL,SCH}. However, for amplitudes that
require regularization the situation is not straightforward and a caution
is necessary. In particular, the LFD dispersion relation between the
energy and the momenta upsets the usual power counting in theories
involving spin.
We give the general formalism first and subsequently give the
results of the detailed calculations.

\subsection{General Formalism}
\label{sect.IV.01}
\noindent
In the LF calculation, integrating over $k^-$ leads to the
time-ordered diagrams shown in Fig.~\ref{Fig1}(b) and (c). 
The pole value $k^-=k^-_{\rm pole}$ is obtained from the on-mass-shell
condition for the corresponding fermion denoted by small circles in 
the diagrams (b) and (c). The diagram (b)
is called the valence part. Depending on the helicity combination
there may be a second contribution, that for $q^+ > 0$ would be 
the diagram (c) called
the nonvalence part, but in our kinematics reduces to the zero-mode
contribution which is the limit $\delta\to 0$ of the nonvalence part.
One can directly calculate the trace term by substitution of $k^-_{\rm
pole}$ in $S^\mu_{h'h}$.

The zero-mode contribution is found using the following identity
\begin{equation}
 \not\!p + m_f = (\not\!{p_{\rm on}} + m_f) + \frac{1}{2}\gamma^+(p^-
- p^-_{\rm on}), 
\label{eq.IV.020}
\end{equation} 
where the subscript (on) denotes the on-mass-shell ($p^2=m^2_f$) fermion
momentum, {\it i.e.}, $p^-=p^-_{\rm on}=(m^2_f+\vec{p}^{\,2}_\perp)/p^+$.
\noindent
Then the trace term $T^+_{h'h}$ of the fermion propagators
in Eq.~(\ref{eq.III.010}) is given by
\begin{eqnarray}
 (T^+_{h'h})_{CP^+} = (T^+_{h'h})_{\rm val} + (T^+_{h'h})_{\rm zm},
\label{eq.IV.030}
\end{eqnarray}
where $(T^+_{h'h})_{\rm val}$ is obtained by substitution of $k^-=
\frac{\vec{k}^{\,2}_\perp +m^2_2}{x p^+}$  
in Eq.~(\ref{eq.III.030}) and $(T^+_{h'h})_{\rm zm}$ is
given by
\begin{eqnarray}
(T^+_{h'h})_{\rm zm} &=&
\frac{1}{2}(k^- - k^-_{\rm on}){\rm Tr} [\not\!\ep^{*}_{h'}\gamma^+
\not\!\ep_{h}
(\not\! k-\not\!p+m_1)\gamma^+ (\not\! k-\not\!p'+ m_1)],
\label{eq.IV.040}
\end{eqnarray} 
where $k^- = p'^- + \frac{m^2_1 + (\vec{k}_\perp -
\vec{p^\prime}_\perp)^2}{k^+-p'^+}$.
As we shall show below, the LF {\em valence} contribution comes
exclusively from the {\em on-mass-shell} propagating part, and the
{\it zero-mode} (if it exists) from the {\it instantaneous} part given
by Eq.~(\ref{eq.IV.040}).

Using the reference frame (See Eq.~(\ref{eq.II.090})) with the LF
gauge, we obtain for the trace terms $(T^+_{h'h})_{\rm val}$ and
$(T^+_{h'h})_{\rm zm}$ the expressions:
\begin{eqnarray}
 (T^+_{++})_{\rm val}&=& \frac{4 p^+}{x}
 \biggl[x^2 m^2_1 +(1-x)^2 m^2_2 + (2x^2-2x + 1)
 \left(\vec{k}^2_\perp - i x Q k_y - \frac{x^2}{4}Q^2\right) \biggr],
\nonumber\\
 (T^+_{+0})_{\rm val}&=&
 \frac{\sqrt{2}p^+}{x M_W} \left\{(2k_x-2ik_y - Qx)(2x-1)
 \left[(k_x + Q x/2)^2 + k^2_y + x(1-x)M^2_W \right] \right.
\nonumber \\
 & & \left. \quad\quad\quad
 + (2 k_x - 2 i k_y + Qx)\left[x m^2_1 -(1-x) m^2_2\right]
 \right\},
\nonumber\\
 (T^+_{+-})_{\rm val}&=& 8(1-x)p^+\left\{(k_x - ik_y)^2 - Q^2 
x^2/4\right\},
\nonumber\\
 (T^+_{00})_{\rm val}&=& \frac{4p^+}{xM^2_W}
 \left\{
 (k^2_x + k^2_y -Q^2 x^2/4)(m^2_1 + m^2_2) + m^2_1 m^2_2
 + \right.
\nonumber\\
 & & \left.
 \left[(k_x +Qx/2)^2 + k^2_y + x(1-x) M^2_W \right]
 \left[(k_x -Qx/2)^2 + k^2_y + x(1-x) M^2_W \right]
 \right\},
\nonumber \\
\label{eq.IV.050}
\end{eqnarray}
and
\begin{eqnarray}
 (T^+_{++})_{\rm zm}&=& 4(p^+)^2 (k^- - k^-_{\rm on})(1-x)^2,
\nonumber\\
(T^+_{+-})_{\rm zm}&=&0,
\nonumber\\
 (T^+_{+0})_{\rm zm}&=& \frac{4(p^+)^2}{\sqrt{2}M_W}(k^- - k^-_{\rm on})
 (1-x) [- k_x+ik_y -Qx/2],
\nonumber\\
 (T^+_{00})_{\rm zm}&=&
 \frac{4(p^+)^2}{M^2_W}(k^- - k^-_{\rm on})
 [\vec{k}^{\,2}_\perp -Q^2x^2/4 + m^2_1],
\label{eq.IV.060}
\end{eqnarray}
where $x=k^+/p^+$ and 
$k^-_{\rm \rm on} = \frac{\vec{k}^{\,2}_\perp +m^2_2}{x p^+}$. 
Since $(k^- - k^-_{\rm on})$ is factored out in $(T^+_{h'h})_{\rm zm}$,
we shall denote the rest by $R^+_{h'h}$,{\it i.e.}, 
$(T^+_{h'h})_{\rm zm}=(k^- - k^-_{\rm on})R^+_{h'h}$.

\subsubsection{Valence contribution} 
\label{sect.IV.1}
\noindent
The valence contribution to the current matrix elements is given by
\begin{eqnarray}
 (G^{+}_{h'h})_{\rm val} &=& \frac{- g^2 Q_f}{(2\pi)^3}
\int^{1}_{0}\frac{dx}{x(1-x)^2}\int d^2 k_\perp
\left(T^+_{h'h}\right)_{\rm val}
\nonumber \\
 & &  \times\frac{1}{M^2_W + Q^2/4 -(\vec{k}^{\,2}_\perp + m^2_2)/x -
 ((\vec{k}_\perp - \vec{p}_\perp)^2 + m^2_1)/(1-x)}
\nonumber \\
 & &  \times\frac{1}{M^2_W + Q^2/4 -(\vec{k}^{\,2}_\perp + m^2_2)/x -
 ((\vec{k}_\perp - \vec{p}'_\perp)^2 + m^2_1)/(1-x)}.
\label{eq.IV.070}
\end{eqnarray}
Here, we aim at performing the integral over $\vec{k}_\perp$
analytically, so we proceed as in the covariant case by introducing a
Feynman parameter to obtain in an obvious notation
\begin{equation}
 \frac{1}{D_1 D_2} = \int^1_0 \frac{dy}{[y D_1 + (1-y)D_2]^2},
\label{eq.IV.080}
\end{equation}
where
\begin{eqnarray}
 D & =  & y D_1 + (1-y) D_2
\nonumber \\
 & = & M^2_W + Q^2/4
 -\frac{\vec{k}^{\,2}_\perp + m^2_2}{x} -
 \frac{\vec{k}^{\,2}_\perp + Q^2/4 + m^2_1}{1-x} + (1-2y)\frac{k_x Q}{1-x}.
\nonumber \\
\label{eq.IV.090}
\end{eqnarray}
By completing the square in $k_x$ and performing the following 
shift in the integration variables
\begin{equation}
 k'_x = k_x -\frac{1-2y}{2} Q x, \quad k'_y = k_y,
\label{eq.IV.100}
\end{equation}
one obtains a denominator that is symmetric in $k'_x$ and $k'_y$:
\begin{equation}
 D =  M^2_W + Q^2/4 - \frac{1}{x(1-x)} 
 \{\vec{k}^{\,\prime 2}_\perp + x m^2_1 + (1-x) m^2_2 +
x(1-x(1-2y)^2Q^2/4) \} .
\label{eq.IV.110}
\end{equation}
After performing the same shift in the trace, one can neglect the
terms that are odd in the components of $\vec{k}^{\,\prime}_\perp$ as 
their
contributions vanish because of symmetry. We denote the shifted trace
by $S^+_{h'h}(\vec{k}^{\,\prime}_\perp)_{\rm val}$.

The final result can now be written in a succinct form
\begin{eqnarray}
 (G^{+}_{h'h})_{\rm val} &=& \frac{- g^2 Q_f}{(2\pi)^3}
 \int^{1}_{0} d x x \,\int^1_0 dy\,\int d^2 k_\perp
  \frac{S^+_{h'h}(\vec{k}_\perp)_{\rm val}}
 {(\vec{k}^{2}_\perp + M^2_W D_{\rm LF})^2},
\label{G-val}
\end{eqnarray}
where the integration variable $\vec{k}^{\,\prime}_\perp$ is changed to
$\vec{k}_\perp$.
The function $D_{\rm LF}$ is
\begin{equation}
 D_{\rm LF} = x \Delta + (1-x) {\cal E} - x(1-x) + 4 x^2 y(1-y) 
\eta.
\label{eq.IV.120}
\end{equation}
Before writing down the traces for the different helicity combinations, we
do the zero-mode first and show that their contributions, if any remain, 
can be written in a form similar to the valence part.

\subsubsection{Zero-mode contribution}
\label{sect.IV.2}
\noindent
The zero-mode is the contribution from the nonvalence part in the limit
$q^+ \to 0$~\cite{BCJ02}. In our kinematics, it is equal to the limit
$\delta\to 0$ of
\begin{eqnarray}
 (G^{+}_{h'h})_{\rm zm} &=& \frac{-g^2 Q_f}{(2\pi)^3 p^+}
 \int^{1+\delta}_{1} \frac{d x}{x(x-1) (x-1 -\delta)} \,\int d^2 k_\perp
\nonumber \\
 & &  \times\frac{R^+_{h'h}}
 {
\frac{\delta(M^2_W + Q^2/4)}{1+\delta}
 + \frac{(\vec{k}_\perp-\vec{p}^{\,\prime}_\perp)^2 + m^2_1}{1+\delta-x}
 + \frac{(\vec{k}_\perp-\vec{p}_\perp)^2 + m^2_1}{x-1}
 },
\label{eq.IV.140}
\end{eqnarray}
where the factor $(k^- - k^-_{\rm on})$ in the trace term 
$(T^+_{h'h})_{\rm zm} = (k^- - k^-_{\rm on}) R^+_{h'h}$ has been cancelled out by the same 
factor in one of the LF energy denominators.
With the substitutions
\begin{equation}
x = 1+ \delta z, \quad k_x = k'_x - \frac{1-x+\delta/2}{\delta} =
 k'_x - \frac{1-2z}{2} Q, \quad k_y = k'_y,
\label{eq.IV.150}
\end{equation}
and taking the limit we obtain
\begin{equation}
 (G^{+}_{h'h})_{\rm zm} = \frac{g^2 Q_f}{(2\pi)^3 p^+}
 \int^1_0 dz\, \int  d^2 k'_\perp
 \frac{ R^+_{h'h}(\vec{k}^{\,\prime}_\perp)}
 {\vec{k}^{\,\prime 2}_\perp + m^2_1 + z(1-z) Q^2}.
\label{eq.IV.160}
\end{equation}
Note here that $(G^+_{h'h})_{\rm zm}$ vanishes if $R^+_{h'h}$(or $(T^+_{h'h})_{\rm zm}$)
carries the factor $(1-x)^n (n \ge 1)$ because $1-x = -\delta z$ goes to zero
as $\delta \rightarrow 0$. This is the case for $(T^+_{++})_{\rm zm}$ and 
$(T^+_{+0})_{\rm zm}$ as shown in Eq.~(\ref{eq.IV.060}). Thus, the nonvanishing 
zero-mode contribution can occur only in $(G^+_{00})_{\rm zm}$.
A straightforward analysis shows that $G^+_{00}$ has indeed a zero-mode.
The shifted trace in the limit $\delta \rightarrow 0$ yields
\begin{equation}
 R^+_{00}(\vec{k}^{\,\prime}_\perp) =
 \frac{4(p^+)^2}{M^2_W}[\vec{k}^{\,\prime 2}_\perp +m^2_1 -z(1-z)Q^2].
\label{eq.IV.170}
\end{equation}
Consequently, changing the integration variables $z$ and
$\vec{k}^{\prime}_{\perp}$ to $x$ and $\vec{k}_{\perp}$, we find for
the zero-mode contribution to $G^+_{00}$:
\begin{equation}
 (G^+_{00})_{\rm zm} = \frac{4 g^2 Q_f p^+}{(2\pi)^3}
 \int^1_0 dx\, \int  d^2 k_\perp
 \frac{\vec{k}^2_\perp + m^2_1 - x(1-x) Q^2} 
 {\vec{k}^2_\perp + m^2_1 + x(1-x) Q^2}.
\label{eq.IV.180}
\end{equation}

\subsection{Individual Current Matrix Elements}
\label{sect.IV.02}
\noindent
Here we give the detailed results of individual current matrix elements. 

\subsubsection{$G^+_{++}$}
\label{sect.IV.02.01}
\noindent
From the procedure illustrated in the previous subsection~\ref{sect.IV.01},
we find 
\begin{equation}
 \left( S^+_{++}(\vec{k}_\perp) \right)_{\rm val}  
 = \frac{4p^+}{x}\left(N^1_{++} \vec{k}^2_\perp + M^2_W N^0_{++} \right),
\label{eq.IV.190}
\end{equation}
where the numerator functions are given by
\begin{eqnarray}
 M^2_W N^0_{++} & = & m^2_1 x^2 + m^2_2 (1-x)^2 - (2x^2 - 2x +1)y(1-y)x^2 Q^2,
\nonumber \\
 N^1_{++} & = & 2x^2 - 2x +1.
\label{eq.IV.190prime}
\end{eqnarray}
Thus, we get 
\begin{eqnarray}
 (G^+_{++})_{\rm val} & = &
 -\frac{4g^2 Q_f  p^+}{(2\pi)^3} \int^1_0 dx \; \int^1_0 dy\; \int d^2 
k_\perp\;
  \frac {\vec{k}^{\, 2}_\perp  N^1_{++} + M^2_W N^0_{++}} 
{[\vec{k}^{\, 2}_\perp + M^2_W D_{\rm LF}]^2}.
\label{eq.IV.200}
\end{eqnarray}
As explained in subsection~\ref{sect.IV.01}, 
$(G^+_{++})_{\rm zm}$ vanishes.

\subsubsection{$G^+_{+0}$}
\label{sect.IV.02.02}
\noindent
Following the same procedure, we find
\begin{equation}
 \left( S^+_{+0}(\vec{k}_\perp) \right)_{\rm val}  =  
 \frac{2\surd 2p^+ Q}{M_W}\left(N^1_{+0}\vec{k}^2_\perp + M^2_W N^0_{+0}
\right),
\label{eq.IV.210}
\end{equation}
where
\begin{eqnarray}
 M^2_W N^0_{+0} & = & (1-y)(xm^2_1 -(1-x) m^2_2) -(2x-1)x(1-x)yM^2_W -
 (2x-1)x^2y(1-y)^2 Q^2,
\nonumber \\
 N^1_{+0} & = & (2x-1)(1-2y).
\label{eq.IV.210prime}
\end{eqnarray}
Owing to the fact that $N^1_{+0}$ is antisymmetric 
and $D_{\rm LF}$ is symmetric under the exchange of $y$ and $1-y$ 
on the interval $(0,1)$, the first part of $(G^+_{+0})_{\rm val}$,
involving $N^1_{+0}$, vanishes and only the second part remains.
Consequently
\begin{eqnarray}
 (G^+_{+0})_{\rm val} & = &
 -\frac{2\sqrt{2} g^2 Q_f p^+ Q}{(2\pi)^3 M_W} \int^1_0 dx \,x \; \int^1_0 
dy\;
 \int d^2 k_\perp
 \frac {M^2_W N^0_{+0}} {[\vec{k}^{\, 2}_\perp + M^2_W D_{\rm LF}]^2}.
\label{eq.IV.220}
\end{eqnarray}
The zero-mode contribution $(G^+_{+0})_{\rm zm}$ vanishes.
         
\subsubsection{$G^+_{00}$}
\label{sect.IV.02.03}
\noindent
For the $h=h'=0$ case, we find
\begin{eqnarray}
 \left( S^+_{00}(\vec{k}_\perp) \right)_{\rm val} & = & 
 \frac{4p^+}{M^2_W x}\left(\vec{k}^4_\perp
 + M^2_W N^1_{00}\vec{k}^2_\perp + M^4_W N^0_{00} \right),
\nonumber \\
 M^4_W N^0_{00} & = & [x(1-x)M^2_W + x^2 y^2 Q^2][x(1-x)M^2_W + x^2(1-y)^2 Q^2]
 + m^2_1 m^2_2 -(m^2_1 + m^2_2)x^2 y(1-y) Q^2,
\nonumber \\
 M^2_W N^1_{00} & = & 2x(1-x)M^2_W + m^2_1+ m^2_2 + x^2(2y-1)^2 Q^2,
\nonumber 
\label{eq.IV.230}
\end{eqnarray}
\begin{eqnarray}
 (G^+_{00})_{\rm val} & = &
 -\frac{g^2 Q_f \,4 p^+}{(2\pi)^3 \,M^2_W} \int^1_0 dx \; \int^1_0 dy\;
 \int d^2 k_\perp
  \frac{\vec{k}^{\, 4}_\perp + \vec{k}^{\, 2}_\perp M^2_W N^1_{00}
 + M^4_W N^0_{00}(x,y,\eta)}
 {[\vec{k}^{\, 2}_\perp + M^2_W D_{\rm LF}]^2},
\nonumber \\
 \left( G^+_{00}\right)_{\rm zm} & = &
 \frac{g^2 Q_f \, 4p^+}{(2\pi)^3\, M^2_W} \int^1_0 dx \; \int d^2 k_\perp 
\;
 \frac{\vec{k}^{\,2}_\perp + m^2_1 - x(1-x) Q^2}
 {\vec{k}^{\,2}_\perp + m^2_1 + x(1-x) Q^2}.
\label{eq.IV.240}
\end{eqnarray}
The angular condition can only be satisfied if the zero-mode
contribution is included, {\it i.e.}, $G^+_{00}=(G^+_{00})_{\rm
val}+(G^{+}_{00})_{\rm zm}$.

\subsubsection{$G^+_{+-}$}
\label{sect.IV.02.04}
\noindent
This is the simplest case because only an absolutely convergent
integral is involved.  Following the procedure described in
subsection~\ref{sect.IV.01}, we find
\begin{equation}
 \left( S^+_{+-}(\vec{k}_\perp)\right)_{\rm val} =
 8(1-x)p^+(k^2_x -k^2_y - y(1-y)x^2 Q^2).
\label{eq.IV.+-}
\end{equation}
However, due to the symmetry between $k_x$ and $k_y$ in Eq.~(\ref{G-val}),
we obtain effectively
\begin{equation}
 \left( G^+_{+-}\right)_{\rm val} = 
  \frac{g^2 Q_f \,8 p^+ Q^2}{(2\pi)^3} \int^1_0 dx \; x^3(1-x) 
 \int^1_0 dy\; y(1-y) \int d^2\vec{k}_\perp \;
 \frac{1}{(\vec{k}^{\, 2}_\perp + M^2_W D_{\rm LF})^2}.
\label{eq.IV.260}
\end{equation}
Moreover,
the zero-mode contribution $(G^+_{+-})_{\rm zm}$ is absent because 
$(T^+_{+-})_{\rm zm} = 0$ as shown in Eq.~(\ref{eq.IV.060}).
Using Eqs.~(\ref{eq.II.110}) and (\ref{eq.IV.260}), it is rather trivial
to reproduce $F_3(0)$ in Eq.~(\ref{eq.III.220}).

\subsection{Regularized LFD Results}
\noindent
In this subsection we present the results for DR${}_2$, SMR,
PV${}_1$ and PV${}_2$. The formal aspects of DR${}_2$
are outlined in the Appendix.
In order to facilitate the discussion we define here, as before, the
elements of the amplitudes that correspond to the struck constituents with
masses $m_a$ and $m_b$ and the spectator with mass $m_2$. The physical
situation is $m_a = m_b = m_1$. The regularization involves replacing
these masses by a cut off $\Lambda$ and taking $\Lambda$ to infinity.
The main ideas were discussed in Sect.~\ref{sect.III}, so we limit the
discussion here to the peculiarities of the LFD  calculation.  Note
that $G^+_{+-}$ is defined by an absolutely convergent integral so it
needs no regularization. Thus, we shall not discuss the latter amplitude
in this subsection.

The denominator function is given by
\begin{equation}
 M^2_W D_{\rm LF}(m_a,m_b;m_2) = x(y m^2_b + (1-y)m^2_a) + (1-x) m^2_2
 - x(1-x) M^2_W + x^2 y(1-y) Q^2 .
\label{eq.IV.291}
\end{equation}
In the numerator the squared mass $m_1^2$ must be replaced by $m_a m_b$, as the
dependence on $m_1$ is due to the mass terms in the fermion propagators
and turn out to be equal to the product of masses.

\vspace{2ex}
\noindent
{\em\bf SMR}

\noindent
In the smearing regularization the masses in the numerators are not replaced by
$\Lambda$. Only the denominators are affected. The regulated amplitude is
\begin{eqnarray}
 (G^{+}_{h'h})^{\rm SMR} &=& -\frac{g^2 Q_f}{(2\pi)^3}
 \int^{1}_{0} d x x \,\int^1_0 dy\,\int d^2 k_\perp
  S^+_{h'h}(\vec{k}_\perp) (m_1;m_2)
\nonumber \\
 & &  \times\left[
 \frac{1}{(\vec{k}^{2}_\perp + M^2_W D_{\rm LF}(m_1,m_1;m_2))^2}
 -\frac{1}{(\vec{k}^{2}_\perp + M^2_W D_{\rm 
LF}(\Lambda_1,m_1;m_2))^2}
 \right.
\nonumber \\
 & &  \left. \quad
 -\frac{1}{(\vec{k}^{2}_\perp + M^2_W D_{\rm 
LF}(m_1,\Lambda_1;m_2))^2}  +
 \frac{1}{(\vec{k}^{2}_\perp + M^2_W D_{\rm 
LF}(\Lambda_1,\Lambda_1;m_2))^2}
 \right].
\end{eqnarray}

\vspace{1ex}
\noindent
{\em Regular parts} (nl)\\
The regular parts proportional to $N^0_{h'h}$ give finite contributions upon
$\Lambda_1 \to\infty$ except for the parts that contain no $\Lambda_1$.
The latter ones vanish in this limit.
These contributions can be read off from our DR${}_2$ results given
above:
\begin{equation}
 (G^{+}_{h'h})^{\rm SMR}_{\rm nl} = \frac{-g^2 Q_f}{2(2\pi)^2}
 \int^{1}_{0} d x x \,\int^1_0 dy\;
 \frac{c_{h'h} M^2_W N^0_{h'h}}{M^2_W D_{\rm LF}}.
\end{equation}
The coefficients $c_{h'h}$ are proportionality constants that can be read off
from the shifted traces, viz
\begin{equation}
 c_{++} = \frac{4p^+}{x}, \quad c_{+0} = \frac{2\sqrt{2} p^+ Q}{M_W},
 \quad c_{00} = \frac{4 p^+}{M^2_W x}.
\label{eq.731}
\end{equation}

\vspace{1ex}
\noindent
{\em Logarithmically divergent parts} (ln)\\
These contributions contain the $1/\epsilon$ terms and logarithms
\begin{eqnarray}
 (G^{+}_{h'h})^{\rm SMR}_{\rm ln} & = &
 \frac{-g^2 Q_f }{2(2\pi)^2} \int^1_0 dx x \int^1_0 dy \;c_{h'h} N^1_{h'h}
\nonumber \\
 &&
 \left[
 \left( \frac{1}{\bar{\epsilon}} +
 \ln\left(\frac{4\pi\mu^2}{M^2_W D_{\rm LF}(m_1,m_1;m_2)} \right) \right)
 -\left( \frac{1}{\bar{\epsilon}} +
 \ln\left(\frac{4\pi\mu^2}{M^2_W D_{\rm LF}(\Lambda_1,m_1;m_2)} \right) \right)
 \right. 
\nonumber \\
 && \left.
 - \left( \frac{1}{\bar{\epsilon}} +
 \ln\left(\frac{4\pi\mu^2}{M^2_W D_{\rm LF}(m_1,\Lambda_1;m_2)} \right) \right)
 + \left( \frac{1}{\bar{\epsilon}} +
 \ln\left(\frac{4\pi\mu^2}{M^2_W D_{\rm LF}(\Lambda_1,\Lambda_1;m_2)} \right)
 \right)
 \right] ,
\end{eqnarray}
where $1/\bar{\epsilon} = 1/\epsilon -\gamma -1$.
It is clear that the  terms $1/\bar{\epsilon}$ cancel. Upon taking the
limit $\Lambda_1 \to \infty$ we obtain a finite logarithm as in the
DR${}_4$ case. The logarithms combine into
\begin{equation}
 -\left\{ \ln (M^2_W D_{\rm LF} (m_1,m_1;m_2))
 - \ln (xy(1-y) \Lambda_1^2) \right\}.
\end{equation}
The final result is then the integral over $x$ and $y$ of this term
multiplied with $c_{h'h} N^1_{h'h}$:
\begin{equation}
 (G^{+}_{h'h})^{\rm SMR}_{\rm ln} = \frac{g^2 Q_f}{2(2\pi)^2}
 \int^{1}_{0} d x x \,\int^1_0 dy\;
  c_{h'h} N^1_{h'h} \left\{ \ln M^2_W D_{\rm LF} (m_1,m_1;m_2) -
 \ln (xy(1-y) \Lambda^2_1 \right\}.
\end{equation}

\vspace{1ex}
\noindent
{\em Quadratically divergent parts} (qu)\\
These terms occur in $G^+_{00}$ only. There are two of them, one comes
from the valence part and the other from the zero-mode. We shall give
them explicitly where we discuss the individual cases.

\vspace{3ex}
\noindent
{\em\bf PV${}_1$}

\noindent
In the PV${}_1$ regularization the mass $m_1$ in the numerators as well as
the denominators is replaced by $\Lambda_1$:
\begin{equation}
 (G^{+}_{h'h})^{{\rm PV}_1} = -\frac{g^2 Q_f}{(2\pi)^3}
 \int^{1}_{0} d x x \,\int^1_0 dy\,\int d^2 k_\perp
  \left[\frac{ S^+_{h'h}(\vec{k}_\perp) (m_1;m_2)}
 {(\vec{k}^{2}_\perp + M^2_W D_{\rm LF}(m_1;m_2))^2}
 -\frac{ S^+_{h'h}(\vec{k}_\perp) (\Lambda_1;m_2)}
 {(\vec{k}^{2}_\perp + M^2_W D_{\rm LF}(\Lambda_1;m_2))^2} \right].
\end{equation}

\vspace{1ex}
\noindent
{\em Regular parts} (nl)\\
The regular parts proportional to $N^0_{h'h}$ give a finite contribution
as $\Lambda_1 \to\infty$ except for the part that contains no
$\Lambda_1$.

\vspace{1ex}
\noindent
{\em Logarithmically divergent parts} (ln)\\
These contributions contain the $1/\epsilon$ terms and logarithms
\begin{eqnarray}
 (G^{+}_{h'h})^{{\rm PV}_1}_{\rm ln} & = &
 -\frac{g^2 Q_f}{2(2\pi)^2} \int^1_0 dx x \int^1_0 dy \; c_{h'h}
\nonumber \\
 &&
 \left[ N^1_{h'h}(m_1;m_2)
 \left(\frac{1}{\bar{\epsilon}} +
 \ln\left(\frac{4\pi\mu^2}{M^2_W D_{\rm LF}(m_1;m_2)} \right) \right)
 - N^1_{h'h}(\Lambda_1;m_2) \left( \frac{1}{\bar{\epsilon}} +
 \ln\left(\frac{4\pi\mu^2}{M^2_W D_{\rm LF}(\Lambda_1;m_2)} \right) \right)
 \right]   . \nonumber \\
\end{eqnarray}
The terms $1/\bar{\epsilon}$ cancel if either the integrals over the
numerator functions are independent of the fermion masses or the integrals
of their coefficients over $x$ and $y$ vanish.

\vspace{1ex}
\noindent
{\em Quadratically divergent parts} (qu)\\
These terms occur in $G^+_{00}$ only. We shall give
them explicitly where we discuss the individual cases.

\vspace{3ex}
\noindent
{\em\bf PV${}_2$}

\noindent
In the PV${}_2$ regularization the mass $m_2$ in the numerators as well
as the denominators is replaced by $\Lambda_2$.
The result of this replacement takes exactly the same form as the ones
for PV${}_1$, except for the replacement of $m_2$ in stead of $m_1$.

\subsection{Summary of Individual $G^+_{h^\prime,h}$ Results}
\label{sect.IV.04}
\noindent
Here we list the results of the integration over $\vec{k}_\perp$ for the
matrix elements using the various regularization methods.

\vspace{2ex}
\subsubsection{$G^+_{++}$}
\label{sect.IV.04.01}

\noindent 
{\em\bf DR${}_2$}

\begin{equation}
 (G^+_{++})^{{\rm DR}_2}_{\rm val} =
 -\frac{2g^2 Q_f p^+}{(2\pi)^2}  \int^1_0 dx \int^1_0 dy
 \left\{ N^1_{++} 
 \left[\left(\frac{1}{\epsilon} - \gamma -1 \right) 
 +\ln \left( \frac{4\pi\mu^2}{M^2_W D_{\rm LF}}\right) \right]
 + \frac{M^2_W N^0_{++}}{M^2_W D_{\rm LF}} \right\}.
\end{equation}
The integral multiplying $1/\epsilon - \gamma -1$ is
\begin{equation}
 \int^1_0 dx \int^1_0 dy\; (2x^2 - 2x +1) = \frac{2}{3}.
\end{equation}

\vspace{2ex}
\noindent
{\em\bf SMR}
\begin{equation}
 (G^+_{++})^{\rm SMR}_{\rm val} = -\frac{2 g^2 Q_f p^+}{(2\pi)^2}  
\int^1_0 dx \int^1_0 dy
 \left\{ N^1_{++} \left[
 \ln \left( \frac{xy(1-y)\Lambda^2}{M^2_W D_{\rm LF}}\right) \right]
 + \frac{M^2_W N^0_{++}}{M^2_W D_{\rm LF}} \right\}.
\end{equation}
The integral giving the correction term is
\begin{equation}
 \int^1_0 dx \int^1_0 dy\; (2x^2 - 2x +1)\ln(xy(1-y)) = -\frac{37}{18}.
\label{++smearing}
\end{equation}

\vspace{2ex}
\noindent
{\em\bf PV${}_1$}
\begin{equation}
 (G^+_{++})^{{\rm PV}_1}_{\rm val} =
 -\frac{2 g^2 Q_f p^+}{(2\pi)^2}  \int^1_0 dx \int^1_0 dy
 \left\{ N^1_{++} \left[
 \ln \left( \frac{x\Lambda^2}{M^2_W D_{\rm LF}}\right) \right]
 + \frac{M^2_W N^0_{++}}{M^2_W D_{\rm LF}} -x \right\}.
\end{equation}
The integrals giving the correction terms are
\begin{equation}
 \int^1_0 dx \int^1_0 dy \; x = \frac{1}{2}, \quad
 \int^1_0 dx \int^1_0 dy\; (2x^2 - 2x +1)\ln x = -\frac{13}{18}.
\end{equation}

\vspace{2ex}
\noindent
{\em\bf PV${}_2$}
\begin{equation}
 (G^+_{++})^{{\rm PV}_2}_{\rm val} =
 -\frac{2 g^2 Q_f p^+}{(2\pi)^2}  \int^1_0 dx \int^1_0 dy
 \left\{ N^1_{++} \left[
 \ln \left( \frac{(1-x)\Lambda^2}{M^2_W D_{\rm LF}}\right) \right]
 + \frac{M^2_W N^0_{++}}{M^2_W D_{\rm LF}} -(1-x) \right\}.
\end{equation}
The integrals giving the correction terms are
\begin{equation}
 \int^1_0 dx \int^1_0 dy \; (1-x) = \frac{1}{2}, \quad
 \int^1_0 dx \int^1_0 dy\; (2x^2 - 2x +1)\ln (1-x) = -\frac{13}{18}.
\end{equation}
Apparently, the corrections are the same for both Pauli-Villars
regularizations.

\vspace{2ex}
\subsubsection{$G^+_{+0}$}
\label{sect.IV.04.02}

\noindent 
{\em\bf DR${}_2$}
\begin{equation}
 (G^+_{+0})^{{\rm DR}_2}_{\rm val} =
 -\frac{\sqrt{2} g^2 Q_f p^+ Q}{(2\pi)^2 M_W}  \int^1_0 dx x \int^1_0 dy\;
  \frac{M^2_W N^0_{+0}}{M^2_W D_{\rm LF}}.
\end{equation}

\vspace{2ex}
\noindent
{\em\bf SMR}
\begin{equation}
 (G^+_{+0})^{\rm SMR}_{\rm val} =
 -\frac{\sqrt{2} g^2 Q_f p^+ Q}{(2\pi)^2 M_W}  \int^1_0 dx x \int^1_0 dy\;
 \frac{M^2_W N^0_{+0}}{M^2_W D_{\rm LF}}.
\end{equation}

\vspace{2ex}
\noindent
{\em\bf PV${}_1$}
\begin{equation}
 (G^+_{+0})^{{\rm PV}_1}_{\rm val} =
 -\frac{\sqrt{2} g^2 Q_f p^+ Q}{(2\pi)^2 M_W}  \int^1_0 dx x \int^1_0 dy\;
 \left[
  \frac{M^2_W N^0_{+0}}{M^2_W D_{\rm LF}} - (1-y) \right]. 
\label{PV1+0}
\end{equation}
The integral giving the correction term is
\begin{equation}
 \int^1_0 dx x \int^1_0 dy \; (1-y) = \frac{1}{4}. \quad
\label{+0PV1}
\end{equation}

\vspace{2ex}
\noindent
{\em\bf PV${}_2$}
\begin{equation}
 (G^+_{+0})^{{\rm PV}_2}_{\rm val} =
 -\frac{\sqrt{2} g^2 Q_f p^+ Q}{(2\pi)^2 M_W}  \int^1_0 dx x \int^1_0 dy\;
\left[ \frac{M^2_W N^0_{+0}}{M^2_W D_{\rm LF}} + (1-y) \right]. 
\label{PV2+0}
\end{equation}
The integral giving the correction term is the same as in the case PV${}_1$.

\subsubsection{$G^+_{00}$}
\label{sect.IV.04.03}

\noindent 
{\em\bf DR${}_2$}
\begin{eqnarray}
 (G^+_{00})_{\rm val} & = &
 -\frac{2 g^2 Q_f \, p^+}{(2\pi)^2 \,M^2_W} \int^1_0 dx \; \int^1_0 dy\; 
 \left\{ -M^2_W D_{\rm LF} 
  \left[
 \frac{2}{\epsilon} - 2\gamma -1 + 
 2 \ln\left(\frac{4\pi\mu^2}{M^2_W D_{\rm LF}} \right)
  \right]
 \right.
\nonumber \\
 & & \quad\quad\quad\quad\quad\quad +  \left. M^2_W N^1_{00}
  \left[
 \frac{1}{\epsilon} - \gamma -1 +
  \ln\left(\frac{4\pi\mu^2}{M^2_W D_{\rm LF}} \right)
  \right]
  + \frac{M^4_W N^0_{00}}{M^2_W D_{\rm LF}}
 \right\},
\label{eq.IV.240a}
\end{eqnarray}
%
\begin{equation}
 \left( G^+_{00}\right)_{\rm zm} = 
  \frac{2 g^2 Q_f \, p^+}{(2\pi)^2\, M^2_W} \int^1_0 dx \;
  2 x(1-x) Q^2 \left[\frac{1}{\epsilon} - \gamma +
 \ln \left( \frac{4\pi\mu^2}{m^2_1 + x(1-x) Q^2} \right)\right].
\label{eq.IV.250a}
\end{equation}
Dimensional regularization removes the part that is quadratically
divergent.  This does not mean that one cannot recover this term. The
terms in the integrals given above that contain the factor $M^2_W
D_{\rm LF}$ in the numerator correspond to the quadratic divergence.
If we now apply for instance PVR, we shall
find a contribution proportional to $\Lambda^2$ in the limit $\Lambda
\to \infty$, which signals the occurrence of the quadratic divergence.
In the formula written above this contribution is concealed because we
have gathered the contribution from the quadratic and log divergences
together.

\vspace{2ex}
\noindent
{\em\bf SMR}
\begin{eqnarray}
 (G^+_{00})^{\rm SMR}_{\rm val} & = &
 -\frac{2 g^2 Q_f \,p^+}{(2\pi)^2 \,M^2_W}
   \int^1_0 dx \; \int^1_0 dy\; 
 \left[2 M^2_W D_{\rm LF}
 \ln \left(\frac{M^2_W D_{\rm LF}}{xy(1-y)(\Lambda^2_1 - m^2_1)}\right)
 \right.
\nonumber \\
 & & \left.
 \quad\quad\quad\quad\quad\quad\quad\quad   - M^2_W N^1_{00}
 \ln \left(\frac{M^2_W D_{\rm LF}}{xy(1-y)(\Lambda^2_1 - m^2_1)}\right)
 + \frac{M^4_W N^0_{00}}{M^2_W D_{\rm LF}} \right],
\\
 \left( G^+_{00}\right)^{\rm SMR}_{\rm zm} & = &
  -\frac{2 g^2 Q_f \,p^+}{(2\pi)^2\, M^2_W} 
 \int^1_0 dx \; 2 x(1-x) Q^2
 \ln \left(\frac{m^2_1 + x(1-x) Q^2}{x(1-x)(\Lambda^2_1 -m^2_1)}\right).
\label{eq.IV.250b}
\end{eqnarray}
The quadratice divergences are cancelled seperately in the valence part as well as in the 
zero-mode part.

\vspace{2ex}
\noindent
{\em\bf PV${}_1$}
\begin{eqnarray}
 (G^+_{00})^{{\rm PV}_1}_{\rm val} & = &
 -\frac{2 g^2 Q_f \,p^+}{(2\pi)^2 \,M^2_W}
 \int^1_0 dx \; \int^1_0 dy\; 
 \left\{
 \left[
 2 M^2_W D_{\rm LF} \ln\left(\frac{M^2_W D_{\rm LF}}{x(\Lambda^2_1 - m^2_1)}
 \right)
- M^2_W N^1_{00} \ln\left(\frac{M^2_W D_{\rm LF}}{x(\Lambda^2_1 - m^2_1)}
 \right)
  + \frac{M^4_W N^0_{00}}{M^2_W D_{\rm LF}}\right]
 \right.
\nonumber \\
 & & \quad\quad\quad\quad\quad\quad \left.
 +  2 M^2_W D_{\rm LF} + \frac{M^2_W D_{\rm LF} -m^2_2 + x^2 y(1-y) Q^2}{x} 
 \right\} ,
\\
 \left( G^+_{00}\right)^{{\rm PV}_1}_{\rm zm} & = &
 -\frac{2 g^2 Q_f p^+}{(2\pi)^2\, M^2_W}
 \int^1_0 dx \;
 2x(1-x) Q^2 \ln \left(\frac{m^2_1 +x(1-x) Q^2}{\Lambda^2_1+x(1-x) Q^2}\right).
\label{eq.IV.250c}
\end{eqnarray}
Again, in PV${}_1$ there is no quadratic divergence.

\vspace{2ex}
\noindent
{\em\bf PV${}_2$}
\begin{eqnarray}
 (G^+_{00})^{{\rm PV}_2}_{\rm val} & = &
 -\frac{2 g^2 Q_f \,p^+}{(2\pi)^2 \,M^2_W}
 \int^1_0 dx \; \int^1_0 dy\;
 \left\{
 \left[
 2 M^2_W D_{\rm LF} \ln\left(\frac{M^2_W D_{\rm LF}}{(1-x)(\Lambda^2_1 - m^2_1)}
 \right)
- M^2_W N^1_{00} \ln\left(\frac{M^2_W D_{\rm LF}}{(1-x)(\Lambda^2_1 - m^2_1)}
 \right)
 \right.\right.
\nonumber \\
 & & \quad\quad\quad\quad\quad\quad \left.\left.
  + \frac{M^4_W N^0_{00}}{M^2_W D_{\rm LF}}\right]
 + 2 M^2_W D_{\rm LF} +\frac{M^2_W D_{\rm LF}-m^2_1 + x^2 y(1-y)Q^2 }{1-x}
 \right\},
\\
 \left( G^+_{00}\right)^{{\rm PV}_2}_{\rm zm} & = & 0 .
\label{eq.IV.250d}
\end{eqnarray}
Apparently, in PV${}_2$ there is no quadratic divergence either.
However, one should note that the zero-mode contribution is removed in
PV${}_2$ by design ({\it i.e.}, artificially). As a consequence of this
artificial removal, $(G^+_{00})^{\rm PV_2}_{\rm val}$ contains the
singular
$x$-integration which makes the calculation in PV${}_2$ impossible.
This is a remarkable result because it shows that if the required
zero-mode contribution is artificially removed then the calculation
cannot be handled in LFD, {\it i.e.}, ``the theory blows up!".

\subsection{Physical quantities}
\noindent
The physical quantities computed in DR${}_2$ using
$F^{+0}_2 + 2 F_1 = (G^+_{+-} + G^+_{+0}/\sqrt{2\eta})/p^+$ are given
by 
\begin{eqnarray}
 F_2(q^2) + 2 F_1(q^2) & = & -\frac{g^2 Q_f}{(2\pi)^2} \int^1_0 dx \;
 \int^1_0 dy\;
 \frac{x[2x^3 -3x^2 + (1+\Delta +\epsilon)x -\epsilon
 -4x^2(3-2x)y(1-y)\eta]}{D_{\rm LF}},
 \nonumber \\
 F_3(q^2) & = & -\frac{g^2 Q_f}{4 \pi^2} \int^1_0 dx \;
 \int^1_0 dy\; \frac{8x^3(1-x) y(1-y)}{D_{\rm LF}}.  
\label{eq.IV.290} 
\end{eqnarray} 
Here, $F_3$ at $q^2=0$ is exactly identical to Eq.~(\ref{eq.III.220})
as we have already noted in subsection~\ref{sect.IV.02.04}.
Furthermore, for any $q^2$, we see that $G^+_{+-}$ involves the
integration of a function of $x$ and $y$ times $1/M^2_W D_{\rm LF}$ and
thus it reduces to the DR${}_2$ value upon taking the limit $\Lambda
\to \infty$ for any of the regularization methods considered.  
Consequently, the predicted value of $F_3(q^2)$ is always the
same regardless of the regularization method. 

However, the combination $F_2(0) + 2 F_1(0)$ does not coincide with the
covariant one in Eq.~(\ref{eq.III.220}), the difference being the integral
\begin{equation}
 A_\kappa =\frac{g^2 Q_f}{(2\pi)^2} \int^1_0 dx \; x(1-x). 
\label{eq.IV.300}
\end{equation}
In fact, we find for any $q^2$
\begin{equation}
F^{+0}_2 + 2 F_1 = (F_2 + 2 F_1)^{\rm DR_4} + \frac{g^2 Q_f}{4\pi^2}\int^1_0 dx x(1-x),
\end{equation}
where we denote the result in Eq.~(\ref{eq.III.201}) as $(F_2 + 2
F_1)^{\rm DR_4}$.  We attribute this fermion-mass-independent
difference to the vector anomaly because it is associated with the
conditionally convergent integral.  The vector anomaly was also
observed in Ref.~\cite{DKMT1} as a difference between the direct
channel and the side-wise channel in manifestly covariant DR${}_4$
calculations. In LFD, it is more interesting to note that this
fermion-mass-independent difference (or vector anomaly) depends on the
choice of helicity amplitudes in computing the same physical quantity.
Using DR${}_2$ but involving $G^+_{00}$, {\it i.e.}, $F^{00}_2+2F_1 =
\{(1+2\eta)G^+_{++}-G^+_{00}+(1+4\eta)G^+_{+-}\}/{4p^+\eta}$, we find
\begin{eqnarray}
F^{00}_2 + 2 F_1 - (F_2 + 2 F_1)_{\rm DR_4} & = &
-\frac{g^2 Q_f}{4\pi^2}\left(\frac{1}{2\eta}\right)\left[\int^1_0 dx \int^1_0 dy
\left\{1+ 2\eta((3-4y(1-y))x^2-2x+1)\right\} - \frac{2}{3}(1+2\eta) 
\right]
\nonumber \\
& = & -\frac{g^2 Q_f}{4\pi^2}\left(\frac{1}{2\eta}\right)\left(\frac{1}{3} +
\frac{2\eta}{9}\right).
\end{eqnarray}
Here, the difference depends on the value of $\eta$ but again is
independent of the fermion mass. Thus, we find
that the vector anomaly in LFD breaks the Lorentz symmetry, {\it i.e.},
$F^{+0}_2 \neq F^{00}_2$.

However, a natural way to remove the vector anomaly in any formulation
(manifestly covariant or LF) is to impose the anomaly-free condition
$\sum_f Q_f = 0$ as in the SM.  This anomaly-free condition restores
the Lorentz symmetry. Now, let's
consider different regularization methods in LFD.

\vspace{2ex}

\noindent
{\em\bf SMR}\\
\noindent
For $G^+_{+0}$, as argued below Eq.~(\ref{eq.IV.210prime}), the
correction term to the DR${}_2$ value vanishes in the limit $\Lambda
\rightarrow \infty$. Thus, the SMR result for $F^{+0}_2 + 2 F_1$ is
identical to the DR${}_2$ result, so the difference with the manifestly
covariant calculation using DR${}_4$ is
\begin{equation}
 (F^{+0}_2 + 2 F_1)^{\rm SMR} - (F_2 + 2 F_1)^{\rm DR_4} =
 \frac{g^2 Q_f}{4\pi^2} \int^1_0 dx x(1-x) =
 \frac{1}{6} \; \frac{g^2 Q_f}{4\pi^2}.
\end{equation}
For $G^+_{00}$, the calculation is highly nontrivial because it
involves not only the zero-mode contribution but also the correction
terms to the DR${}_2$ values both in the valence part and the zero-mode
part do not vanish in the limit $\Lambda \rightarrow \infty$.  However,
the calculation of $F^{00}_2 + 2 F_1$ involves also $G^+_{++}$ which
also deviates from the DR${}_2$ value as shown in
Eq.~(\ref{++smearing}).  It is really remarkable that all of these
deviations conspire to give exactly identical result between $(F^{00}_2
+ 2 F_1)^{\rm SMR}$ and $(F^{+0} + 2 F_1)^{\rm SMR}$, viz.
\begin{eqnarray}
(F^{00}_2 + 2 F_1)^{\rm SMR} - (F_2 + 2 F_1)^{\rm DR_4} &=&
-\frac{g^2 Q_f}{4\pi^2}\left(\frac{1}{2\eta}\right)\left[\int^1_0 dx 
\int^1_0 dy
\left\{1+ 2\eta((3-4y(1-y))x^2-2x+1)\right\}\right. 
\nonumber \\
&+& \left.
 \left(\frac{20\eta}{9}-\frac{19}{18}\right)_{(G^+_{00} {\rm ~correction})}
 +(1+2\eta)\left(-\frac{37}{18} \right)_{(G^+_{++} {\rm ~correction})}\right] 
\nonumber \\
&=&  \frac{1}{6} \; \frac{g^2 Q_f}{4\pi^2}.
\end{eqnarray}

Thus, the SMR results for the physical quantity are identical
regardless of the choice in the helicity amplitudes. Moreover, one
should note that the SMR results in LFD are identical to the
SMR result from the manifestly covariant calculation, {\it i.e.}, as
shown in Eq.~(\ref{eq.III.240})

\begin{equation}
(F_2 + 2 F_1)^{\rm SMR}_{\rm cov} - (F_2 + 2 F_1)^{\rm DR_4} 
= \frac{1}{6} \;\frac{g^2 Q_f}{4\pi^2},
\end{equation}
so that 
\begin{equation}
(F^{+0}_2 + 2 F_1)^{\rm SMR} 
= (F^{00}_2 + 2 F_1)^{\rm SMR} 
= (F_2 + 2 F_1)^{\rm SMR}_{\rm cov}. 
\end{equation}
This shows that all the SMR results are absolutely convergent and
restore the Lorentz symmetry completely. However, the
fermion-mass-independent difference between the SMR result and the
manifestly covariant DR${}_4$ result clearly exists. As we show below,
the same conclusion is obtained also in the calculations with the PVR
supporting the existence of the vector anomaly further.

\vspace{2ex}

\noindent
{\em\bf PV${}_1$}\\
\noindent
For $G^+_{+0}$, as shown in Eq.~(\ref{+0PV1}), the correction term to
the DR${}_2$ value does not vanish. Thus, the PV${}_1$ result for
$F^{+0}_2 + 2 F_1$ is modified from the DR${}_2$ result, {\it i.e.},
\begin{eqnarray}
(F^{+0}_2 + 2 F_1)^{\rm PV_1} - (F_2 + 2F_1)^{\rm DR_4}
& = & \frac{g^2 Q_f}{4\pi^2}\left[\left(\frac{1}{6}\right)_{({\rm from~DR_2})}
+ \left(\frac{1}{2}\right)_{(G^+_{+0} {\rm ~correction})}\right]
\nonumber \\
& = & \frac{2}{3} \; \frac{g^2 Q_f}{4\pi^2}.
\end{eqnarray}
The calculation involving $G^+_{00}$ and $G^+_{++}$ is highly
nontrivial as explained already in the SMR case. However, it is
again remarkable that all the correction terms conspire to give exactly
identical result regardless of the choice of helicity amplitudes, {\it
i.e.},
\begin{eqnarray}
(F^{00}_2 + 2 F_1)^{\rm PV_1} - (F_2 + 2 F_1)^{\rm DR_4}
& = & -\frac{g^2 Q_f}{4\pi^2}\left(\frac{1}{2\eta}\right)
 \left[
 \int^1_0 dx \int^1_0 dy \; \{1+ 2\eta((3-4y(1-y))x^2-2x+1)\} 
 \right.
\nonumber \\
 & & 
 \left.
 - \left(\frac{4\eta}{9}-\frac{2}{9}\right)_{(G^+_{00}{\rm ~correction})} 
 + (1+2\eta)
 \left(-\frac{13}{18} - \frac{1}{2}\right)_{(G^+_{++} {\rm ~correction})}
 \right]
\nonumber \\
 & = & \frac{2}{3} \; \frac{g^2 Q_f}{4\pi^2}.
\end{eqnarray}
Also, we note that the PV${}_1$ results in LFD are identical to the
PV${}_1$ result from the manifestly covariant calculation because, as
shown in Eq.~(\ref{eq.III.250}),
\begin{equation}
(F_2 + 2 F_1)^{\rm PV_1}_{\rm cov} - (F_2 + 2 F_1)^{\rm DR_4}
 = \frac{2}{3} \; \frac{g^2 Q_f}{4\pi^2},
\end{equation}
so that
\begin{equation}
(F^{+0}_2 + 2 F_1)^{\rm PV_1} 
= (F^{00}_2 + 2 F_1)^{\rm PV_1} 
= (F_2 + 2 F_1)^{\rm PV_1}_{\rm cov}.
\end{equation}
Thus, the same conclusion for the PV${}_1$ results arises as in the case of 
SMR; {\it i.e.}, the PV${}_1$ results are absolutely convergent and
restore completely
the Lorentz symmetry. However, the fermion-mass-independent difference
between the 
PV${}_1$ results and the manifestly covariant DR${}_4$ result persists
further supporting
the existence of the vector anomaly.

\vspace{2ex}

\noindent
{\em\bf PV${}_2$}\\
\noindent
The correction term for $G^+_{+0}$ in the PV${}_2$ case has the same
magnitude but opposite sign from that in the PV${}_1$ case (see
Eqs.~(\ref{PV1+0}) and (\ref{PV2+0})). Thus, the PV${}_2$ result for
$F^{+0}_2 + 2 F_1$ is given by
\begin{eqnarray}
(F^{+0}_2 + 2 F_1)^{\rm PV_2} -(F_2 + 2 F_1)^{\rm DR_4} &=&
\frac{g^2 Q_f}{4\pi^2}\left[\left(\frac{1}{6} \right)_{\rm from~DR_2} - 
 \left(\frac{1}{2} \right)_{G^+_{+0} {\rm ~correction}}\right]
\nonumber \\ 
&=& -\frac{1}{3} \; \frac{g^2 Q_f}{4\pi^2}.
\label{PV2F+0}
\end{eqnarray}
In the PV${}_2$ case, however, the calculation involving $G^+_{00}$
cannot be completed because the zero-mode contribution is
artificially removed and causes a singular $x$-integration in the
valence part as we have discussed in subsection~\ref{sect.IV.04.03}.
This assures that the zero-mode contribution in $G^+_{00}$ is essential
to make the LFD calculation not only correct but also possible. Thus, the 
only way to avoid the zero-mode contribution in the form factor calculation
is to make a judicious choice of the helicity amplitudes. Without
involving $G^+_{00}$, we found the result in Eq.~(\ref{PV2F+0}) for
instance.
Note here again that $(F^{+0}_2 + 2 F_1)^{\rm PV_2}$ is identical to
the PV${}_2$ result from the manifestly covariant calculation,{\it
i.e.}, as shown in Eq.~(\ref{eq.III.260}),
\begin{equation}
(F_2 + 2 F_1)^{\rm PV_2}_{\rm cov} - (F_2 + 2 F_1)^{\rm DR_4}
= -\frac{1}{3} \; \frac{g^2 Q_f}{4\pi^2}.
\end{equation}
Thus, the PV${}_2$ results are uniquely obtained regardless of the formulation
(LFD or manifestly covariant calculation) and yield another
fermion-mass-independent
deviation from the DR${}_4$ result supporting the existence of
the vector anomaly.

\section{Discussion and Conclusion}
\label{sect.VI}
\noindent 
Besides the orthodox dimensional
regularization, denoted by DR${}_4$, of the manifestly covariant
formulation in the Weinberg-Salam theory, we considered 
other regularizations,
smearing (SMR) and Pauli-Villars (PVR) regularization in this paper.  
In the spirit of
the original paper by `t Hooft and Veltman~\cite{tHV72}, we also studied
a variant of dimensional regularization that extends the dimensionality
of space in the transverse directions only, which we called DR${}_2$.
In all cases, the corrections to the CP-even photon-$W^\pm$ vertex
given by the triangle diagram could be clearly separated into divergent
parts and finite ones. Of the physical observables, the charge,
magnetic dipole moment, and electric quadrupole moment, only the charge
needs renormalization. The other moments must be finite and predicted
by the theory. The quadrupole moment turns out to be expressed in terms
of a convergent integral, both in the manifestly covariant formulation
and the LFD formulation. Moreover, the integral defining the quadrupole
moment is the same for all regularization methods, as it should be.

In QED, where there is no anomaly cancellation, the situation is quite
different.  For instance, the triangle diagram associated with the
fermion-photon vertex correction has two parts, $F^{\rm QED}_1$ and
$F^{\rm QED}_2$.  The former is given by a divergent integral and the
latter by a convergent one. This can be compared to the Weinberg-Salam
case where $F^{W^\pm}_1$ and $F^{W^\pm}_2$ are given by divergent
integrals while $F^{W^\pm}_3$ is expressed in terms of a convergent
one.  The situation for the charge and quadrupole form factors is
similar to $F^{\rm QED}_1$ and $F^{\rm QED}_2$. The former must be
renormalized while the latter is finite and unaffected by the
regularization method used.  The physical magnetic dipole form factor
$F^{W^\pm}_2 + 2F^{W^\pm}_1$ is, however, given by a conditionally
convergent integral.  That is the reason why the magnetic form factor
needs a special care.

In view of the fact that the charge must be renormalized, the precise
form of its finite part is less interesting than the magnetic moment.
In the manifestly covariant formulation, the latter turns out to depend
on the regularization method used.  As shown in this work,
the fermion-mass-independent finite differences exist among the $F_2 +
2F_1$ results
with different regularization methods. They result from the different
ways of handling the bad infinities and thus can be regarded as a
symptom of the vector anomaly existing in the triangle-fermion-loop
considered in our calculation.  However, this has no consequences for
the predictive power of the Weinberg-Salam theory, as the differences
between the various results is a constant multiple of the charge $Q_f$
which is independent of the fermion masses. So, if a sum over the
fermionic contributions to the magnetic moment is taken, the
contribution of these anomalous parts is proportional to the sum over
$Q_f$ which vanishes in all three generations of the SM.
This is the celebrated property of anomaly cancellation, which
apparently saves the day.  According to the non-renormalization
theorem\cite{non-renormalization}, the cancellation of the fermion
one-loop anomaly implies anomaly cancellation to all orders in
perturbation theory.  We should also note that the consistency with the
Furry's theorem~\cite{Furry} is assured in the vector anomaly because
the charge factor $Q_f$ cancels out between the fermion and antifermion
contributions for the neutral spin-1 particles.

In LFD, the physical quantities are obtained from a linear
combination of helicity matrix elements of the vertex operator. 
Specifically, in DR${}_2$, the form factor $F_2 + 2 F_1$
turns out to depend on whether the matrix element $G^+_{00}$ is
involved or not.  In either case, a discrepancy with the results
obtained using DR${}_4$ is found, which, however, is again a constant
multiple of the charge $Q_f$ which is independent of the fermion
masses. We think this is quite relevant to our former
discussion~\cite{BJ1} on the fact that the common belief of the
equivalence between the manifestly covariant formulation and the
light-front Hamiltonian formulation is not always justified. In the
explicit example of a 1+1 dimensional exactly solvable model
calculation, we have shown the existence of an end-point singularity in
the $J^-$ current matrix element which spoils the equivalence between
the LFD result and the manifestly covariant result. Later, we have
shown the recovery of the equivalence using the SMR which removes the
end-point singularity~\cite{BCJ01}.  The difference between the
DR${}_4$ and DR${}_2$ results is a reminiscence of this inequivalence
between Euclidian and Minkowski calculations as also shown in
Ref.~\cite{BJ1}. What we find in this work is more significant in the
sense that the present LFD calculation involved entirely the
so-called good current $J^+$ and the difference is not just a
singularity but a finite measurable quantity. Moreover, if $G^+_{00}$
is used, the difference with DR${}_4$ does depend on the momentum
transfer squared. Without having any protection from an explicit cutoff
parameter $\Lambda$, the dimensional regularizations DR${}_4$ and
DR${}_2$ seem to reveal the real difference between the Euclidian and
Minkowski formulations.  Thus, in the dimensional regularizations, the
vector anomaly caused by the bad infinity not only spoils
the common belief in the equivalence but also violates the Lorentz
symmetry. Unless the anomaly-free condition is strictly fulfilled, the
differences among $(F_2 + 2F_1)_{{\rm DR}_4}$, $(F_2 +2F_1)^{0+}_{{\rm
DR}_2}$ and $(F_2 +2F_1)^{00}_{{\rm DR}_2}$ would remain.

However, using the SMR and PVR with an explicit cutoff parameter
$\Lambda$, we can show that the difference between the Euclidian and
Minkowski formulations is removed just like the recovery
of the equivalence we have shown in Ref.~\cite{BCJ01}. 
Although $\Lambda \rightarrow\infty$ is taken in our end results, 
the regularization with an
explicit $\Lambda$ assures the absolute convergence of the
loop calculation. Thus, not only the equivalence between the Euclidian
and Minkowski formulations is recovered but also the Lorentz symmetry
(or the angular condition) is satisfied. The symptom of a vector anomaly
nevertheless appears as finite constant differences in $F_2
+ 2F_1$ dependent on the regularization methods. Although their
appearances are rather different from the dimensional regularization
cases, these finite differences are again fermion-mass-independent and
thus removed under the usual anomaly-free condition in the SM.

We should note that the zero-mode contribution in $G^+_{00}$ is crucial
to obtain these results. In particular, in the case of PV${}_2$ where
the zero-mode is artificially removed, the theory blows up in the sense
that the singular $x$-integration in the valence part of $G^+_{00}$
makes the calculation in PV${}_2$ impossible. When the zero-mode
contribution is taken into account as shown in SMR and PV${}_1$, the
quadratic divergences are removed in the LFD calculations and the
angular condition is satisfied.  These results make it clear that
light-front quantization is different from the manifestly covariant
formulation using a Wick rotation to Euclidian momenta and dimensional
regularization. 

According to our findings in this work, we can discard the possibility
of having a point-particle model for vector mesons, {\it e.g.}, the
$\rho^\pm$ mesons.  If such a model is used, not only will two of the
three form factors ($F_1,\, F_2,\, F_3$)  be infinite, but also there
will occur finite differences depending on the regularization methods
in the $F_2 + 2F_1$ prediction for the $\rho^\pm$ mesons. These
differences in the prediction of a physical quantity depending on the
regularization method cannot be removed as the $\rho^\pm$ mesons are
bound-states of a quark and an antiquark even if the charge
renormalization is applied. This situation is certainly not acceptable
in a viable model prediction.  Therefore, any reasonable model for the
composite $\rho^\pm$ systems should have a finite hadronic size which
may correspond to a finite cutoff parameter $\Lambda$ ({\it e.g.}, in
SMR, see also Ref.~\cite{BCJ02}). As we have shown in this work,
introducing a cutoff parameter $\Lambda$ assures the equivalence
between the Euclidian and Minkowski formulations and Lorentz symmetry
(satifying the angular condition). Moreover, the finite value of
$\Lambda$ in a particular regularization scheme would correspond to an
input parameter in a particular hadronic model.  Although we cannot
tell whether such a model can be derived from the first principles of
QCD or not, we can at least show that the latter model with a finite
$\Lambda$ can pass the fitness test illustrated in this work, while the
former point-particle (or $\Lambda \rightarrow \infty$ limit) model
cannot.  The similar concept of a fitness test can be applied to the
deuteron for models with point nucleons.  Thus, our findings in this
work may provide a bottom-up fitness test not only to the LFD
calculations but also to the theory itself, whether it is any extension
of the Standard Model or an effective field theoretic model for
composite systems.  Further investigations on the CP-odd form factors,
the relation to the Ward identity, the sidewise channel and bound-state
applications may deserve further considerations.

\acknowledgements
\noindent
We thank Stan Brodsky for suggesting the calculation of the zero-mode
contribution in the case of $W^\pm$ gauge bosons. We also thank Al
Mueller and Gary McCartor for illuminating discussions.  CRJ thanks the
hospitality of Piet Mulders and the discussion with Daniel Boer at the
Vrije Universiteit where this work was completed.  
BLGB wants to thank the faculty and  staff of the Department of Physics
of North Carolina State University for their hospitality during his stay
in Raleigh where this work was started. This work was
supported in part by the grant from the U.S. Department of Energy
(DE-FG02-96ER40947) and the National Science Foundation (INT-9906384).

\appendix

\section{Mathematical Details}
\label{sect.A}
\noindent
For momentum integrals we go over from four to $D$-dimensions. 

\subsection{Euclidian space: DR${}_4$}
\noindent
The pertinent integration may be written as 
\begin{equation}
 \int\frac{d^4 k}{(2\pi)^4} f(k) \to \mu^{4-D} \int \frac{d^D
k}{(2\pi)^D} f(k).
\label{eq.A.010}
\end{equation}
We include the factor $\mu^{4-D}$ to keep the dimension of the integral
the same as in four dimensions~\cite{PT84}.  For divergent integrals we
write $D = 4 - 2\epsilon$.  Next the limit $\epsilon\to 0$ is taken. Then
we find the well-known formula

\begin{equation}
 I^r_s = 
 \mu^{4-D}\,\int
 \frac{d^Dk}{(2 \pi)^D} \frac{(k^2)^r}{(k^2 - C^2)^s}  = \mu^{2\epsilon}
 \frac{i(-1)^{(r+s)}}{(4\pi)^{D/2}}
 \frac{\Gamma(r+D/2)\Gamma(s-r-D/2)}{\Gamma(D/2)\Gamma(s)}
 \left(\frac{1}{C^2}\right)^{s-r-D/2}.
\label{eq.A.020a}
\end{equation}
The factor $i(-1)^{(r+s)}$ originates  in the Wick rotation from
Minkowski space to Euclidian space.
In particular, we find
\begin{eqnarray}
 I^0_3 = \mu^{4-D}\, \int \frac{d^Dk}{(2 \pi)^D} \frac{1}{(k^2 - C^2)^3}
 & = & \frac{-i}{2 (4\pi)^2} \frac{1}{C^2}, 
\nonumber \\
 I^1_3 = \mu^{4-D}\,\int \frac{d^Dk}{(2 \pi)^D} \frac{k^2}{(k^2 - C^2)^3} & = &
 \frac{i}{(4\pi)^{2-\epsilon}}
 \frac{\Gamma(3-\ep)\Gamma(\ep)}{\Gamma(2-\ep)\Gamma(3)}
 \left(\frac{\mu^2}{C^2}\right)^\epsilon 
\nonumber \\
 & \to &  \frac{i}{(4\pi)^2}
 \left(\frac{1}{\ep} - \gamma -\frac{1}{2} \right)
  + \frac{i}{(4\pi)^2} \ln\left(\frac{4\pi\mu^2}{C^2}\right), 
\nonumber\\
 I^2_3 = \mu^{4-D}\,\int \frac{d^Dk}{(2 \pi)^D} \frac{k^4}{(k^2 - C^2)^3} & = &
 \frac{-i\mu^{2\epsilon}}{(4\pi)^{2-\epsilon}}
 \frac{\Gamma(4-\ep)\Gamma(-1+\epsilon)}{\Gamma(2-\epsilon)\Gamma(3)}
 \left(\frac{1}{C^2}\right)^{-1 +\epsilon }
\nonumber \\
 & \to &  \frac{i\,C^2}{(4\pi)^2}
 \left(\frac{3}{\ep} - 3\gamma +\frac{1}{2} \right)
  + 3\frac{i\,C^2}{(4\pi)^2} \ln\left(\frac{4\pi\mu^2}{C^2}\right). \nonumber\\
\label{eq.A.020} 
\end{eqnarray}

For the Gamma function, we use
\begin{eqnarray}
 \Gamma(z+1) & = & z \Gamma(z)\nonumber \\
 \Gamma(\epsilon) & = & \frac{1}{\epsilon} - \gamma + {\cal O}(\epsilon).
\label{eq.A.020b}
\end{eqnarray}

\subsection{Minkowski space: DR${}_2$}
\noindent
For the 2D integrals over $\vec{k}_\perp$ we need not perform a Wick
rotation. We can use a similar procedure as before, but need to change
it a little bit.  Now we take $D = 2 - 2\epsilon$, defining the integral
\begin{equation}
 \bar{I}^r_s = 
 \mu^{2-D}\,\int
 \frac{d^Dk}{(2 \pi)^D} \frac{(k^2)^r}{(k^2 + C^2)^s}  = \mu^{2\epsilon}
 \frac{1}{(4\pi)^{D/2}}
 \frac{\Gamma(r+D/2)\Gamma(s-r-D/2)}{\Gamma(D/2)\Gamma(s)}
 \left(\frac{1}{C^2}\right)^{s-r-D/2}.
\label{eq.A.020c}
\end{equation}
Because some subtleties may occur, we give two examples: $\bar{I}^1_1$
and $\bar{I}^2_2$ which contain the quadratic divergences:
\begin{equation}
 \bar{I}^1_1 = \frac{\mu^{2\epsilon}}{(4\pi)^{D/2}} \frac{1}{(C^2)^{-D/2}}
 \frac{\Gamma(1+D/2)\Gamma(-D/2)}{\Gamma(D/2)\Gamma(1)}
\label{eq.A.20c}
\end{equation}
and
\begin{equation}
 \bar{I}^2_2 = \frac{\mu^{2\epsilon}}{(4\pi)^{D/2}} \frac{1}{(C^2)^{-D/2}}
 \frac{\Gamma(2+D/2)\Gamma(-D/2)}{\Gamma(D/2)\Gamma(2)}.
\label{eq.A.20ca}
\end{equation}
Writing $D = 2 - 2\epsilon$ it is tempting to neglect the $\epsilon$'s in
the arguments of the Gamma functions in case it is of the form
$1-\epsilon$, $2-\epsilon$ etc. Since doing so causes a change in the
finite part of the calculation, we write
\begin{equation}
 \bar{I}^1_1 =\frac{C^2}{4\pi} 
 \left( \frac{4\pi\mu^2}{C^2}\right)^\epsilon \;
 \frac{\Gamma(2-\epsilon)\Gamma(-1+\epsilon)}{\Gamma(1-\epsilon)\Gamma(1)}
\label{eq.A.20d}
\end{equation}
and
\begin{equation}
 \bar{I}^2_2 =\frac{C^2}{4\pi} 
 \left( \frac{4\pi\mu^2}{C^2}\right)^\epsilon \;
 \frac{\Gamma(3-\epsilon)\Gamma(-1+\epsilon)}{\Gamma(1-\epsilon)\Gamma(2)}.
\label{eq.A.20da}
\end{equation}
Using the relations given above, we find the results
\begin{equation}
 \bar{I}^1_1 = -\frac{C^2}{4\pi}
 \left[ \frac{1}{\epsilon} -  \gamma +
 \ln \left( \frac{4\pi\mu^2}{C^2} \right) \right]
\label{eq.A.20ea}
\end{equation}
and
\begin{equation}
 \bar{I}^2_2 = -\frac{C^2}{4\pi}
 \left[ \frac{2}{\epsilon} - 2 \gamma -1 +
 2 \ln \left( \frac{4\pi\mu^2}{C^2} \right) \right].
\label{eq.A.20e}
\end{equation}
In summary, the integrals we needed are given by
\begin{eqnarray}
 \bar{I}^0_0 & = & 0, 
\nonumber \\
 \bar{I}^0_1 & = &  \frac{1}{4\pi} \left[
 \frac{1}{\epsilon} - \gamma + \ln\left(\frac{4\pi\mu^2}{C^2} \right)\right]
\nonumber \\
 \bar{I}^1_1 & = & -\frac{C^2}{4\pi}
 \left[ \frac{1}{\epsilon} -  \gamma +
 \ln \left( \frac{4\pi\mu^2}{C^2} \right) \right]
\nonumber \\
 \bar{I}^0_2 &=& \frac{1}{4\pi C^2}
\nonumber \\
 \bar{I}^1_2 & = &  \frac{1}{4\pi} \left[
 \frac{1}{\epsilon} - \gamma - 1 + \ln\left(\frac{4\pi\mu^2}{C^2}
\right)\right]
\nonumber \\
 \bar{I}^2_2 &=&  -\frac{C^2}{4\pi}
 \left[ \frac{2}{\epsilon} - 2 \gamma -1 +
 2 \ln \left( \frac{4\pi\mu^2}{C^2} \right) \right].
\label{eq.A.20g}
\end{eqnarray}

\end{document}